# Anomalies in Light Scattering

Alex Krasnok[1*], Denis Baranov[2], Huanan Li[1], Mohammad-Ali Miri[3], Francesco Monticone[4], and Andrea Alú[1*]

[1]*Advanced Science Research Center, City University of New York, New York, NY 10031, USA*

[2]*Department of Physics, Chalmers University of Technology, 412 96 Gothenburg, Sweden*

[3]*Department of Physics, Queens College of the City University of New York, Queens, New York 11367, USA*

[4]*School of Electrical and Computer Engineering, Cornell University, Ithaca, NY 14853, USA*

E-mail: akrasnok@gc.cuny.edu, aalu@gc.cuny.edu

## Abstract

Scattering of electromagnetic waves lies at the heart of most experimental techniques over nearly the entire electromagnetic spectrum, ranging from radio waves to optics and X-rays. Hence, deep insight into the basics of scattering theory and understanding the peculiar features of electromagnetic scattering is necessary for the correct interpretation of experimental data and an understanding of the underlying physics. Recently, a broad spectrum of exceptional scattering phenomena attainable in suitably engineered structures has been predicted and demonstrated. Examples include bound states in the continuum, exceptional points in $\mathcal{PT}$-symmetrical non-Hermitian systems, coherent perfect absorption, virtual perfect absorption, nontrivial lasing, non-radiating sources, and others. In this paper, we establish a unified description of such exotic scattering phenomena and show that the origin of all these effects can be traced back to the properties of poles and zeros of the underlying scattering matrix. We provide insights on how managing these special points in the complex frequency plane provides a powerful approach to tailor unusual scattering regimes.

## Introduction

The wave nature of light has been recognized since the pioneering works of Huygens, who published his *Treatise on Light* in 1690[1]. This hypothesis was supported by the following experiments and theoretical developments by Young, Ampere, Fresnel, Faraday, and others. Almost two centuries after Huygens' work, Maxwell developed a complete mathematical description of electromagnetic radiation published in *A Treatise on Electricity and Magnetism*[2]. This formulation, known as Maxwell's equations for more than 150 years has been the basis for understanding light propagation, radiation, and scattering in various systems, ranging from planar surfaces and lenses to sophisticated nanophotonic structures.

In most macroscopic structures, light undergoes simple scattering (i.e., a change in the propagation direction of the incident wave) processes, such as specular reflection and transmission, absorption, diffuse scattering, focusing. With the advent of metamaterials and nanophotonics, researchers have begun investigating the interaction of light with more complicated structures[3–8]. They have been shown to exhibit unusual scattering responses, including super-focusing and super-resolution[9,10], negative refraction[11], extraordinary transmission, localization in random media[3], radiation control[12], and even slowdown of propagating electromagnetic waves[13–15]. In these phenomena, unusual spatial distributions of the electromagnetic fields stem from the response of individual or collective resonant structures, as in the case of plasmonic nanoparticles or photonic crystals, which is not achieved in non-resonant optical structures.

Recent studies have shown that even more exotic and fundamentally different scattering responses can emerge in some engineered nanophotonic structures. Theoretical predictions and subsequent observations of exceptionally small or large scattering from nanoparticles, coherent perfect absorption and PT-symmetric phase transitions, non-trivial interplay of gain and absorption, embedded eigenstates and virtual absorption are some of the phenomena, recently emerged in the scientific literature, which have proven that interaction of light with matter can go far beyond our classical representation of light scattering, and may become highly counter-intuitive. Understanding the general picture of these seemingly different phenomena, their physical mechanisms, and the deep relations between them is crucial to exploit the technological potential of these effects and further develop engineered optical structures with novel functionalities. This review paper aims to establish this general picture and provide analytical insights into the connection between various extreme *scattering anomalies*.

To begin with, we formulate the general problem of light scattering and restrict the range of possible systems considered in this paper. The classical statement of the electromagnetic scattering problem considers the interaction of the incident field with an object localized in one, two or three dimensions. Maxwell's equations describing the spatial and temporal evolution of electric $\mathbf{E}(\mathbf{r},t)$ and magnetic $\mathbf{H}(\mathbf{r},t)$ fields for an arbitrary system are[16]

$$\nabla \times \mathbf{E}(\mathbf{r},t) = -\partial_t \mathbf{B}(\mathbf{r},t), \quad \nabla \cdot \mathbf{D}(\mathbf{r},t) = \rho(\mathbf{r},t), \tag{1a}$$

$$\nabla \times \mathbf{H}(\mathbf{r},t) = \mathbf{J}(\mathbf{r},t) + \partial_t \mathbf{D}(\mathbf{r},t), \quad \nabla \cdot \mathbf{B}(\mathbf{r},t) = 0, \tag{1b}$$

where $\mathbf{D} = \varepsilon_0 \mathbf{E} + \mathbf{P} = \varepsilon_0 \hat{\varepsilon} \mathbf{E}$, $\mathbf{B} = \mu_0(\mathbf{H} + \mathbf{M}) = \mu_0 \hat{\mu} \mathbf{H}$, $\rho$ and $\mathbf{J}$ stand for free charge density and current, $c$ is the speed of light in vacuum, $\mathbf{P}$ and $\mathbf{M}$ are the electric and magnetic polarizations, respectively, $\varepsilon_0$ is the dielectric constant, $\hat{\varepsilon}$ and $\hat{\mu}$ are the (relative) permittivity and permeability (tensors and/or integral operators, in general). A general problem of light scattering consists in finding the solutions of Eq. (1) with the following boundary conditions: $\mathbf{n} \times (\mathbf{E}_2 - \mathbf{E}_1) = 0$, $\mathbf{n} \cdot (\mathbf{D}_2 - \mathbf{D}_1) = \sigma$, $\mathbf{n} \times (\mathbf{H}_2 - \mathbf{H}_1) = \mathbf{K}$, and $\mathbf{n} \cdot (\mathbf{B}_2 - \mathbf{B}_1) = 0$, where $\mathbf{n}$ is the normal vector from medium "1" to medium "2", and $\varsigma / \mathsf{K}$ is surface (free) charge/current density.

In this paper, we restrict our consideration to *linear*, *nonmagnetic* ($\hat{\mu} = \hat{I}$), *isotropic* ($\hat{\varepsilon} = \varepsilon\hat{I}$) materials, where $\varepsilon$ can be inhomogeneously distributed and complex-valued and $\hat{I}$ is the identity matrix. The materials are assumed to be uncharged, i.e., free charge density $\rho$ and currents **J** assumed to be zero. We also assume the host medium to be dispersionless. When the scattering problem is linear, we assume a monochromatic excitation, so that all electromagnetic fields involved in the problem depend on time through a factor $\exp(-i\omega t)$.

In the following sections, we show that the scattering ($\hat{S}$) matrix approach offers a powerful toolbox for the description of light scattering in such systems. It turns out that the $\hat{S}$-matrix and its eigenvalues have scattering singularities, *poles* and *zeros*, which correspond to special regimes when scattering from an object becomes infinite or vanishes entirely[17–29], Section (1). Mathematically, these states correspond to Maxwell solutions with only outgoing waves and only incoming waves, respectively. Control over these singularities allows predicting new interesting phenomena and engineering light scattering. In this paper, we focus on perfect absorption (PA), coherent perfect absorption (CPA) and virtual perfect absorption (VPA), Section (2), exceptional points in $\mathcal{PT}$-symmetric non-Hermitian systems, Section (3), bound states in the continuum (BICs), Section (4). Zeros of the scattering amplitude (reflection coefficient, Mie scattering coefficients) have a different physical meaning (no scattered waves, whereas outgoing fields can reside) and give rise to other scattering phenomena, including cloaking, anapole, and tunneling, Section (5). In conclusion, we summarize all considered phenomena in terms of pole and zero analysis in Table (1). We believe that our paper sheds new light on the intimate relations between different exceptional optical scattering effects being currently studied by several groups, and will stimulate further research in this exciting area.

## 1. Theoretical background

### 1.1 Scattering matrix

We assume a linear scatterer is situated in a dielectric medium ($\varepsilon_h$) and described by an inhomogeneously distributed complex-valued relative dielectric permittivity $\varepsilon_s(\omega, \mathbf{r})$. We can define $\Delta\varepsilon(\omega, \mathbf{r}) = \varepsilon_s(\omega, \mathbf{r}) - \varepsilon_h$ and separate the total electric field into the incident $\mathbf{E}_i$ and scattered $\mathbf{E}_s$ components, $\mathbf{E} = \mathbf{E}_i + \mathbf{E}_s$. The total electric field satisfies the Helmholtz equation

$$\nabla \times \nabla \times \mathbf{E}(\omega, \mathbf{r}) - [\Delta\varepsilon(\omega, \mathbf{r}) + \varepsilon_h]\frac{\omega^2}{c^2}\mathbf{E}(\omega, \mathbf{r}) = 0, \qquad (2)$$

which stems from Eqs. (1) in the frequency-domain. This equation is equivalent to the *Fredholm integral equation* for the scattered field

$$\mathbf{E}_s(\mathbf{r}) = \frac{\omega^2}{c^2}\int \hat{G}(\omega, \mathbf{r}, \mathbf{r}')\Delta\varepsilon(\omega, \mathbf{r}')[\mathbf{E}_i(\mathbf{r}') + \mathbf{E}_s(\mathbf{r}')]d^3\mathbf{r}'. \qquad (3)$$

Here, $\hat{G}(\omega, \mathbf{r}, \mathbf{r}')$ is the Green's function of the Helmholtz equation in the host medium[30]. In the operator notation Eq. (3) is written as $\mathbf{E}_s = \hat{L}(\mathbf{E}_i + \mathbf{E}_s)$, where $\hat{L}$ is the integral operator. Formally, the solution of this equation can be written with the use of the *scattering operator* $\hat{S}$:

$$\mathbf{E}_s = \hat{S}\mathbf{E}_i = (\hat{I} - \hat{L})^{-1}\hat{L}\mathbf{E}_i. \tag{4}$$

The above equation is the formal solution of the scattering problem.

To get a deeper understanding of the mechanics underlying various resonant effects, we need to have a closer look at the structure of $\hat{S}$. In practice, it is convenient to represent the incident and scattered waves in a specific *basis of incoming and outgoing waves*, or *channels*, as depicted in Fig. 1(a):

$$\mathbf{E}_i = \sum_p s_p^+ \mathbf{E}_p^+, \quad \mathbf{E}_s = \sum_p s_p^- \mathbf{E}_p^-. \tag{5}$$

The set of amplitudes for the incoming and outgoing waves is denoted as $\mathbf{s}^+ = \{s_1^+, s_2^+, ...\}$ and $\mathbf{s}^- = \{s_1^-, s_2^-, ...\}$, respectively. Hence, in the basis of these channels, Eq. (4) can be rewritten as

$$\mathbf{s}^- = \hat{S}\mathbf{s}^+, \tag{6}$$

where $\hat{S}$ is the *scattering matrix* of the system[31][32]. Essentially, Eq. (6) represents the matrix formulation of the scattering problem: amplitudes of the outgoing waves are linear combinations of the incoming wave amplitudes, and this relationship is established by the scattering matrix, as depicted in Fig. 1(a). Usually, the channels are normalized such that $|s_p^+|^2$ and $|s_p^-|^2$ correspond to the energy flux of the incoming and outgoing waves in a channel $p$.

Using the definition of channels in Eq. (5), we can introduce scattering $S_{pq}$ and reflection $S_{pp} \equiv S_p$ coefficients as

$$S_{pq} = \frac{s_q^-}{s_p^+}, \quad S_p = \frac{s_p^-}{s_p^+}, \tag{7}$$

which describe scattering from a channel $p$ to a channel $q$ and reflection to the same channel $p$, respectively. For example, in the case of extended flat objects, considered in Section (2), channels $p$ and $q$ are simply incident ("1"), reflected ("2"), and transmitted ("3") plane waves. Therefore, Eqs. (7) define the well-known Fresnel's reflection ($r \equiv S_{12} = s_2^-/s_1^+$) and transmission $t \equiv S_{13} = s_3^-/s_1^+$ coefficients. In the case of 2D or 3D scatterers, see Section (5), the channels may be cylindrical or spherical harmonics. Equation (7) allows us to introduce energy conservation relations, $|S_{pq}|^2, |S_p|^2 \leq 1$, which apply to any *passive* (namely, no generation of energy) system. Generally, this statement is not valid for $\mathcal{PT}$-symmetrical systems, discussed in Section (3), since they involve active elements.

The channels introduced in Eq. (5) exist outside the system and represent freely propagating solutions of the wave equation in the absence of the scattering object. The specific choice of channels often depends on the exact geometry of the system. For a planar slab and normal incidence, for example, the basis of normally incident plane waves is the obvious choice, whereas, for a spherical particle, it is convenient to use a basis of vector spherical harmonics.

## 1.2 Hamiltonian approach

It is instructive to introduce the effective Hamiltonian approach, which provides a bottom-up scheme to construct the S-matrix of a system and also highlights the essential features and limitations of the *coupled-mode theory*[33–37] for wave scattering in various photonic structures[34,38–40]. To some extent, this effective Hamiltonian approach can be considered an advanced form of the standard coupled-mode theory, taking into account the dispersive properties of the channels[41]. Let us consider, to be specific, an optical scatterer described by the effective Hermitian Hamiltonian $\hat{H}_0$, which is coupled to $M$ propagating channels labeled by $p = 1, 2, \ldots, M$. In the modal space of the scatterer, the full Hamiltonian $\hat{H}_0$ is given as a $N \times N$ Hermitian matrix with $N$ being the number of supported modes.

The unperturbed portion of $\hat{H}_0$ is typically associated with an isolated ideal scatterer with canonical geometry, for which the modes are analytically solvable. Correspondingly, the diagonal terms of the Hamiltonian $\hat{H}_0$ represent the resonant frequencies of the normal modes in this basis, while the off-diagonal ones denote the coupling between them due to imperfections, such as deformations, defects, etc.[42]. On the other hand, the propagating channel $p$ is associated with the $p$-th waveguide coupling one of the impinging channels to the scatterer.

For waveguides fulfilling the Helmholtz equation, Ref.[41] develops a general treatment of the corresponding Hamiltonian model. Here, for simplicity we consider Coupled-Resonator Optical Waveguides (CROWs) satisfying the Schrodinger-like equation[43]. We set the distance between the nearest-neighbor resonators to be $a_0 = 1$ with dispersion $\omega = 2c_p \cos k_p$ for the $p$-th channel, after suitably fixing the reference frequency, and $c_p < 0$ is the coupling strength between its nearest sites/resonators. Generally, within the channel $p$, the field profile of a monochromatic wave can be written in terms of counter-propagating waves as

$$\psi_p(x,t) = \frac{s_p^+}{\sqrt{\upsilon_{g,p}}} e^{-i\omega t + ik_p x} + \frac{s_p^-}{\sqrt{\upsilon_{g,p}}} e^{-i\omega t - ik_p x}, \tag{8}$$

where the group velocity $v_{g,p} \equiv \partial \omega / \partial k_p = -2c_p \sin k_p$, $k_p \in (0, \pi)$ and the integer $x = -1, -2, \ldots, -\infty$ labels the sites in each waveguide. In Eq. (8), $|s_p^\pm|^2$ are normalized to be the incoming/outgoing flux for each channel $p$.

If $M$ channels at $x = -1$ are coupled through the scatterer with $N$ modes, a real $M \times N$ bare coupling matrix $-\hat{W}$ describes the process. Correspondingly, Schrodinger-like equations describing this setup can be written as

$$i\frac{d}{dt}\mathbf{a}(t) = \hat{H}_0 \mathbf{a}(t) - \hat{W}^T \boldsymbol{\psi}(-1,t), \tag{9a}$$

$$i\frac{d}{dt}\boldsymbol{\psi}(-1,t) = \hat{c}\boldsymbol{\psi}(-2,t) - \hat{W}\mathbf{a}(t), \tag{9b}$$

where $\boldsymbol{\psi}(x,t) = (\psi_1, \psi_2, \cdots, \psi_M)^T$, $\hat{c} = \text{diag}\{c_1, c_2, \cdots c_M\}$ and the vector $\mathbf{a}(t) = (a_1, a_2, \cdots, a_N)^T$ represents the set of mode amplitudes, which are normalized such that $|a_n|^2$ corresponds to the energy in the $n$-th mode. Note that Eq. (9a) can be considered a definition of the Hamiltinian $\hat{H}_0$.

We proceed to derive the scattering matrix $\hat{S}$ based on Eq. (9). First, we derive the coupled-mode equation for the wave scattering from Eq. (9). Essentially setting $\mathbf{a}(t) = \mathbf{a}(0)e^{-i\omega t}$ and plugging Eq. (8) into Eq. (9), we get the coupled-mode equations

$$-i\omega \mathbf{a}(0) = -i(\hat{H} - i\hat{\Gamma})\mathbf{a}(0) - \hat{K}^T \mathbf{s}^+, \tag{10a}$$

$$\mathbf{s}^- = -\mathbf{s}^+ - \hat{K}\mathbf{a}(0), \tag{10b}$$

with the effective Hamiltonian

$$\hat{H} = \hat{H}_0 - \frac{1}{2}\hat{K}^T \cot \hat{k} \hat{K} \tag{11}$$

where $\Gamma = \frac{1}{2}\hat{K}^T \hat{K}$ [44], and $\hat{K} \equiv \hat{c}^{-1}\sqrt{\hat{v}_g}\hat{W}$ is the normalized $M \times N$ coupling matrix, with $\hat{v}_g \equiv$ diag $\{v_{g,1}, v_{g,2}, \cdots, v_{g,M}\}$ being the group-velocity matrix of the channels. The scattering matrix $\hat{S}$ can be obtained from Eq. (10) by eliminating $\mathbf{a}(0)$ and using the definition of $\hat{S}$. Specifically, we have

$$\hat{S} = -\hat{I}_M + i\hat{K}\frac{1}{\omega - \hat{H} + i\hat{\Gamma}}\hat{K}^T, \tag{12}$$

This expression explicitly contains the scattering matrix $\hat{S}$ of an object expressed in terms of its Hermitian Hamiltonian $\hat{H}_0$, coupling operator $\hat{K}^T$, and the incident frequency $\omega$ of the monochromatic wave. Interestingly, the structure of the scattering matrix Eq. (12) is general, *irrespective of the model for the channels*[45,46], and can be extended even more generally starting from the vector Helmholtz equation Eq. (2) (see Ref. [47] and references therein). We also note that although Eq. (12), as well as the coupled-mode theory, is derived in the weak coupling regime, it may be generalized to strongly coupled systems by changing the basis to "dressed states" [39].

Useful distribution information about scattering singularities can be extracted from Eq. (12) by assuming proper ensembles of the random matrix $\hat{H}_0$ based on the preserved symmetry class, such as the Gaussian orthogonal/unitary ensemble (GOE/GUE) for isolated scatterers with/without time-reversal symmetry[48,49]. Nevertheless, the information required for the reconstruction of the scattering matrix $\hat{S}$ based on the Hamiltonian $\hat{H}_0$ and the coupling operator $\hat{K}$ may be overwhelming. Therefore, it is meaningful to develop an approach for the reconstruction of $\hat{S}$ using the minimal available knowledge, say the complex eigenfrequencies and the far-field properties of the *quasinormal modes* (or resonant states) of the system, see Ref.[50–52].

The *quasinormal modes* (QNMs) of photonic structures are defined as the solution of Eq. (9) with *outgoing radiation boundary conditions*[51,53,54], whose completeness has been proven for simple geometries (e.g., spherically symmetric objects) and is a conjecture in other scenarios[52,55–57]. The QNMs are associated with the poles of the scattering matrix in Eq. (12). To see this clearly, setting $\mathbf{s}^+ = 0$ in Eq. (10a), corresponding to outgoing boundary conditions, we get a set of complex eigenfrequencies of the quasinormal modes $\tilde{\omega}_n$ and a corresponding eigenset of QNMs profiles $\tilde{\mathbf{a}}_n(0)$, $n = 1, 2, \cdots, N$. The complex eigenfrequencies $\{\tilde{\omega}_n\}$ turn out to be the poles of the scattering matrix $\hat{S}$ in Eq. (12), i.e., $\det(\tilde{\omega}_n - \hat{H} + i\Gamma) = 0$. Moreover, the asymptotic behavior of the QNMs in the outgoing channels is $\tilde{\mathbf{b}}_n = -\hat{K}\tilde{\mathbf{a}}_n(0)$, which defines the structure of radiation leaking from resonant states into free space.

Next, we can reconstruct the scattering matrix $\hat{S}$ based on its quasinormal mode expansion, following Ref.[50]. It is important to point out the implicit assumption in optical systems that tacitly ignores the frequency-dependence of the Hermitian Hamiltonian $\hat{H}$ and coupling operators $\hat{K}$. This assumption is the analog of the so-called *wideband approximation* in condensed-matter physics[58]. For the reconstruction, the required information can be written compactly as the $N$-dimensional diagonal matrix $\tilde{\Omega} = \text{diag}(\tilde{\omega}_1, \tilde{\omega}_2, \cdots, \tilde{\omega}_N)$ for the complex eigenfrequency spectrum and the $M \times N$ matrix $\tilde{B} = (\tilde{\mathbf{b}}_1, \tilde{\mathbf{b}}_2, \cdots, \tilde{\mathbf{b}}_N)$ for the far-field properties of the QNMs. First, we notice that

$$(\hat{H} - i\hat{\Gamma})\tilde{A} = \tilde{A}\tilde{\Omega}, \tag{13a}$$

$$\tilde{B} = -\hat{K}\tilde{A}, \tag{13b}$$

where the $N \times N$ matrix $\tilde{A} = (\tilde{\mathbf{a}}_1(0), \tilde{\mathbf{a}}_2(0), \cdots, \tilde{\mathbf{a}}_N(0))$ with the columns being the profile of the QNMs inside the scatterer. The above relations allow us to rewrite the S-matrix in Eq. (12) as

$$\hat{S} = -\hat{I}_M + i\tilde{B}\frac{1}{\omega - \tilde{\Omega}}\Lambda^{-1}\tilde{B}^T, \tag{14}$$

where $\Lambda \equiv \tilde{A}^T \tilde{A} = \text{diag}(\lambda_1, \lambda_2, \cdots, \lambda_N)$ is a diagonal matrix, which follows from the *reciprocity of the structure*, i.e., mathematically $(\hat{H} - i\hat{\Gamma})^T = \hat{H} - i\hat{\Gamma}$. Next, we determine the matrix $\Lambda$ (or specifically its diagonal entries $\lambda_N$) from the given information of the quasinormal-modes $\tilde{\Omega}$ and $\tilde{B}$. Since the coupling operator $\hat{K}$ is a real matrix, we have the identity $\tilde{B}^*(Q^*)^{-1}\Lambda = \tilde{B}$, where $Q \equiv \tilde{A}^+ \tilde{A}$ with components $Q_{nn'} = i\frac{\tilde{b}_n^+ \tilde{b}_{n'}}{\tilde{\omega}_n^* - \tilde{\omega}_{n'}}$, which is determined completely from the available information[50]. For convenience, we can define $(\mathbf{x}_1, \mathbf{x}_2, \cdots, \mathbf{x}_N) = \tilde{B}^*(Q^*)^{-1}$ with the vector $\mathbf{x}_n$ as the columns of the latter matrix, which allows us to rewrite the last equation equivalently as $\mathbf{x}_n \lambda_n = \tilde{\mathbf{b}}_n$, where $n = 1, 2, \cdots, N$. Under the present formalism, this last equation is overdetermined but consistent to be used to determine $\lambda_N$. Considering potential numerical or measurement errors, a unique solution of $\mathbf{x}_n \lambda_n = \tilde{\mathbf{b}}_n$ in the least square sense can be given, which reads $\lambda_n = \mathbf{x}_n^+ \tilde{\mathbf{b}}_n / \mathbf{x}_n^+ \mathbf{x}_n$. Finally, the desired quasinormal-mode expansion of the scattering matrix follows[50]

$$\hat{S} = -\hat{I}_M + i\sum_{n=1}^{N} \frac{\mathbf{x}_n^+ \mathbf{x}_n}{\mathbf{x}_n^+ \tilde{\mathbf{b}}_n} \frac{\tilde{\mathbf{b}}_n \tilde{\mathbf{b}}_n^T}{\omega - \tilde{\omega}_n}, \tag{15}$$

where the vector $\mathbf{x}_n$ can be written explicitly as $\mathbf{x}_n = \sum_{n'}(Q_{n'n}^{-1})^* \tilde{\mathbf{b}}_{n'}^*$.

The material gain/loss can be easily incorporated into the previous formalism by setting $\hat{H}_0 \to \hat{H}_0 - i\Gamma_{nr}$ in Eq. (12), where $\Gamma_{nr}$ is a Hermitian decay matrix. When the non-radiative gain/loss is small compared to the frequency of the mode, the diagonal non-radiative gain/loss terms $\Gamma_{nr} = \text{diag}(\gamma_{nr,1}, \gamma_{nr,2}, \cdots, \gamma_{nr,N})$ account for the gain ($\gamma_{nr,n} < 0$) and loss ($\gamma_{nr,n} > 0$) of the modes. Correspondingly, the matrix components of $Q$ in $(\mathbf{x}_1, \mathbf{x}_2, \cdots, \mathbf{x}_N) = \tilde{B}^*(Q^*)^{-1}$ will become $Q_{nn'} = i\frac{\tilde{b}_n^+ \tilde{b}_{n'}}{\tilde{\omega}_n^* - i\gamma_{nr,n} - \tilde{\omega}_{n'} - i\gamma_{nr,n'}}$ [50].

It is also important to stress that the quasinormal-mode expansion Eq. (15) does not depend on the normalization of the far-fields $\tilde{\mathbf{b}}_n$. Also, in the above theoretical exploration, an explicit phase relation is fixed between the input and output fields to eliminate redundant degrees of freedom. In other words, we have $\hat{C} = -\hat{I}_M$ for the direct coupling operator $\hat{C}$ between the incoming and outgoing channels in the absence of the scatterer [38,50]. This choice ensures that the incoming wave will be completely reflected when disconnected from the scatterer. Note that the direct coupling operator accounts for the relative phase of the input and output waves and it is not related to QNMs or poles. Interested readers may resort to Ref.[59] for other choices and relevant discussions.

Finally, it is worthwhile to briefly mention that the Hamiltonian approach for optical systems can be extended to time-modulated systems[60], hold the promise for Floquet scattering engineering[61,62], and may facilitate the study of the reconfigurable control of scattering anomalies like Refs.[63,64]. However, the exploration along this line is beyond the context of this paper.

## 1.3 Poles and zeros of the S-matrix

Generally, the S-matrix $\hat{S}$ can be reduced to its diagonal form $\hat{S}_D = \text{diag}(d_1, d_2, \cdots, d_M)$ after diagonalization. Each eigenvalue $d_m(\omega)$ is associated with its incoming eigenvector $\mathbf{s}_m^+$

$$\hat{S}(\omega)\mathbf{s}_m^+ = d_m(\omega)\mathbf{s}_m^+. \tag{16}$$

In the case of scatterers without material gain/loss, i.e., when the permittivity and permeability are real, the scattering eigenvalues $d_m(\omega)$ for real $\omega$ are unimodular, i.e., $|d_m(\omega)|=1$. This corresponds to a unit circle in the complex domain. This constraint is imposed by the unitarity of the scattering matrix, i.e., $\hat{S}\hat{S}^\dagger = \hat{I}_M$, which is associated with a Hermitian Hamiltonian $\hat{H}_0 = \hat{H}_0^\dagger$ of the scatterer[65,66]. The physical meaning of these eigenvalues becomes apparent when we present them in the exponential form $d_m(\omega) = \exp[2i\delta_m(\omega)]$. Indeed, such representation shows that an incident field corresponding to the eigenvector $\mathbf{s}_m^+$ upon scattering on the system experiences phase shift $\delta_m(\omega)$, corresponding to the incident eigenvector being multiplied by its eigenvalue $d_m$.

When the frequency $\omega$ is analytically continued in the complex plane, the scattering eigenvalues $d_m(\omega)$ can take any values between $\infty$ and 0. These limit values correspond to the poles and zeros of the S-matrix eigenvalues, respectively. The zeros of the S-matrix are different from the "scattering zeros" (absence of scattered waves), as discussed in Section (1.4). The poles refer to the QNMs, as discussed before, while the zeros are associated with vanishing outgoing fields (perfect absorbing regimes), which will be discussed in Section (2). For now, we point out that, for lossless scatterers with $\hat{H}_0 = \hat{H}_0^\dagger$, poles $\omega_{\text{rs},n} = \tilde{\omega}_n$ and zeros $\omega_{\text{pa},n}$ occur in *complex conjugate pairs*, i.e., $\omega_{\text{pa},n} = \omega_{\text{rs},n}^*$. Indeed, the zeros $\omega_{\text{pa},n}$ satisfy the equation $\det(\omega_{\text{pa},n} - \hat{H} - i\Gamma) = 0$. Moreover, after a Hermitian transpose operation, this equation for the zeros becomes $\det(\omega_{\text{pa},n}^* - \hat{H}^\dagger + i\Gamma) = 0$, which, for a Hermitian Hamiltonian $\hat{H}$ is the secular equation for the poles, implying that the poles $\omega_{rs,n} = \omega_{pa,n}^*$ for Hermitian systems, as expected from time-reversal symmetry considerations.

A fundamental property of the scattering matrix eigenvalues $d_n$, which is going to be important for further discussion, is that they can be expanded into a special product *via* the *Weierstrass factorization theorem*[26,67]:

$$d_n(\omega) = A_n \exp(iB_n\omega) \prod_m \frac{\omega - \omega_{pa,m}}{\omega - \omega_{rs,m}}, \qquad (17)$$

where $A_n$ and $B_n$ are constants, and the product is taken over all resonances which match the symmetry of $d_n$. These expressions include only information about the positions of poles $\omega_{rs,m}$ and zeros $\omega_{pa,m}$ in the complex plane. Therefore, the Weierstrass expansion shows that the scattering process is *fully described only by the positions of poles $\omega_{rs,m}$ and zeros $\omega_{pa,m}$*. This fact makes it possible to describe and classify a variety of scattering phenomena in terms of the position of zeros and poles. It is worth noting that Eq. (17) works only for meromorphic functions, when any finite area of the complex frequency plane contains a finite number of poles [52,68,69].

> The scattering process is fully described by the position of poles and zeros. This makes possible to describe, classify, and predict a variety of scattering phenomena.

We note that this technique for scattering calculations based on the S-matrix zeros and poles has been widely used in the study of gratings, especially in the context of Wood's anomalies [23,24,70–75]. These features arise when new diffraction channels open (with an increase in frequency or an increase in the structure period), accompanied by the appearance of a diffracted beam parallel to the surface of the grating. In this paper, we do not discuss Wood's anomalies [76], and we refer the interested readers to several comprehensive works [77–79].

### 1.4 Scattering zeros and poles

Besides poles and zeros of the $\hat{S}$ matrix, we can use poles and zeros of other physical quantities. For example, the reflection coefficient ($R(\omega) = |r(\omega)|^2$) also has singularities in the complex frequency plane. Because a pole originates as an eigenfrequency of a QNM, $R(\omega)$ shares the same poles with the $\hat{S}$ matrix eigenvalues. However, in general the scattering zeros (sometimes called reflectionless scattering modes) do not coincide with the zeros of the $\hat{S}$-matrix, since they correspond to the absence of scattered waves, whereas the outgoing fields in the channels are unperturbed incident waves. In the scenario of scattering by extended objects, a scattering zero on the a real frequency axis corresponds to the well-known tunneling at a Fabry–Pérot resonance (vanishing reflection along with total transmission) [80], tunneling through epsilon-near-zero PT-symmetric layered structures [81,82], and epsilon-near-zero structures [83]. Interestingly, unlike the zeros of the S-matrix, the position of scattering zeros depends on the excitation arrangement. Generally, it depends on signals in all excitation channels, their amplitudes and phases. This provides a great opportunity for coherent all-optical scattering control [84]. For example, two-port coherent excitation can result in coalescencing of a pole and a zero of the scattering coefficient with BIC-like scattering lines in reflection [84].

Scattering zeros of 3D structures play an even more important role because they give rise to a plethora of unusual scattering phenomena, including cloaking, anapole [85], Fano resonance, and coupled-resonator induced transparency [86]. For example, let us consider a small spherical resonator. For analysis of such particle, one can use the Mie scattering theory, which yields, for example, the equation for the scattering cross-section (SCS) [87,88], containing scattering coefficients in the form $c_l = -U_l / (U_l - jV_l)$ [89]. Here, the quantities $U_l^{TE,TM}$ and $V_l^{TE,TM}$ are defined by a combination of spherical Bessel and Neumann functions [90]. These amplitudes have scattering zeros when $U_1 = 0$ giving rise to cloaking [91] or anapoles [85]. Such nonscattering states are of great interest in modern photonics and have been employed for boosting nonlinear effects in dielectric nanoparticles, such as third harmonic generation and four-wave mixing [92–95]. In its turn, the scattering poles are realized when the denominator turns to zero, $U_1^i - jV_1^i = 0$. The pole and zero of the 3D structure can also coalesce at the real axis by tailoring the structure geometry with the realization of the bound state in the continuum (BIC) scattering regime [96].

## 1.5 Q-factor

In scattering problems, QNMs of a system manifest themselves as resonant enhancement of scattering/extinction cross-section, reflection/transmission coefficients, etc. at *real frequencies*. Since the energy leaks out from the scatterer and gets dissipated, each specific quasinormal mode $l$ is characterized by a complex frequency $\omega_l = \text{Re}[\omega_l] + i\text{Im}[\omega_l]$, with the position at the real axis $\text{Re}[\omega_l]$ and a finite lifetime $\tau_l = 1/\gamma_l = -1/\text{Im}[\omega_l]$, where $\gamma_l$ is the *decay rate* and determines the resonance linewidth at its half-maximum[54]. A resonance is usually described by *quality (Q) factor*, which is defined as

$$Q = -\frac{\text{Re}[\omega_l]}{2\text{Im}[\omega_l]}. \tag{18}$$

Factor 2 occurs in the denominator because the energy decays with the decay rate $2\gamma$. In general, the cavity decay rate includes both radiative and absorptive parts $\gamma = \gamma^{rad} + \gamma^{loss}$, such that the quality factor can be expressed as $Q^{-1} = Q_{rad}^{-1} + Q_{loss}^{-1}$ [3].

The Q-factor in the form of Eq. (18) requires an important note. By its definition, the Q-factor defines the decay rate of a resonant state, or the *number of oscillations of the resonant state* *before it decays by a factor of 2.7*. Thus, Eq. (18) is applicable to a single QNM, uncoupled from other resonant nodes of the system (orthogonal to them). Otherwise, turning-off of a monochromatic excitation causes its spectral broadening and inevitable excitation of other (not orthogonal) QNMs.

Interference of different QNMs plays a crucial role in various effects, e.g., Fano-resonances, anapoles, bound states in the continuum (BIC), and others. Hence, although many papers report "high-Q Fano resonances", the definition of Q-factor in these coupled resonance phenomena requires special handling. For example, an *optical Fano resonance* [as well as anapoles, see Section (5)] is caused by interference of radiation from at least two QNMs, bright and dark ones, in the far-field [97,98]. The coupling of these nonorthogonal modes can lead to speeding up the decay rate of the dark mode and, hence, the temporal dynamics are no longer directly related with the Q-factor of the individual modes.

## 1.6 S-matrix symmetry properties

We summarize here the general requirements imposed on the S-matrix by different physical realizations and symmetries.

*Reciprocity* implies that if a signal can travel a certain path, the same signal should be able to travel back with the same transmission coefficient[99–104]. If the system is reciprocal, its scattering operator is symmetric: $\hat{S} = \hat{S}^T$. For example, for a 2-port structure it provides equal transmission in any direction, $t_{12} = t_{21}$. For a 3-port system, this requirement yields, $t_{12} = t_{21}$, $t_{13} = t_{31}$, and so on.

Reciprocity has a great deal of importance in this discussion, because it poses fundamental constraints on the electromagnetic response of a system [35] and on the inverse design of optical devices [105]. However, there are several practical situations (e.g., isolators, filters, full-duplex communication systems) in which it is advantageous to break reciprocity. Reciprocity can be broken ($\hat{S} \neq \hat{S}^T$) *via* the use of magnetized materials [106], optical nonlinearity in spatially asymmetric structures [107], or employing time modulation [104]. The holy grail in this field is a so-called *ideal isolator*, a device allowing signals to propagate outwards from the source and preventing reflections towards the source. This functionality can be generalized to a 3-port system, for which the S-matrix reads as

$$\hat{S} = \begin{pmatrix} 0 & 0 & 1 \\ 1 & 0 & 0 \\ 0 & 1 & 0 \end{pmatrix}. \tag{19}$$

*Photonic topological insulators* [108,109] are artificially designed materials supporting nonreciprocal one-way edge states propagating around sharp bends or other types of defects without backscattering. This fascinating property has been recently connected with the physics of nonreciprocal electromagnetic [110] and acoustic structures [111].

*Hermitian scatterers* with $\hat{H}_0 = \hat{H}_0^\dagger$. By a Hermitian scatterer, we denote a system without *material* gain/*loss*, and thereby we use the fact that *radiation* to the propagating channels does not spoil the flux conservation of the stationary scattering process. However, one has to bear in mind that the Hamiltonian of a scatterer in an open set-up is non-Hermitian with complex-valued QNMs when radiation losses are included[112]. For systems without *material* gain/*loss*, i.e., when the permittivity and permeability are real everywhere, $\hat{S}$ is unitary at the real frequency, $\hat{S}^\dagger \hat{S} = \hat{I}$ and its eigenvalues obey $|d_m(\omega)|^2 = 1$ [see Eq. (16)] corresponding to each scattering channel. For such systems, the frequencies of poles and zeros are restricted to specific parts of the complex frequency plane. QNMs (resonant states) of a passive system must decay in time, due to leakage of the energy from the system into free space. With the time dependence defined through the factor $\exp(-i\omega t)$, the attenuation corresponds to the complex frequencies $\omega_{\text{rs},n}$ in the lower half-plane. Correspondingly, the frequencies of perfectly absorbing modes (S-matrix zeros) $\omega_{\text{pa},n} = \omega_{\text{rs},n}^*$ sit symmetrically in the upper half-plane, Figs. 2(a,b). One exception is a bound state in the continuum (BIC), Fig. 2(b), which is a QNM of the scatterer with a real frequency in the continuum of unbounded radiation modes and can exist in lossless or $\mathcal{PT}$-symmetrical[113–115] structures infinite in at least one direction[116] or for extreme values of parameters[96,117–119], see Section (4).

When the coupling $\hat{K}$ between the scatterer and the channels decreases, the pairs of poles and zeros approach the real axis symmetrically. Finally, in the limit of vanishing coupling, they coincide with the eigenfrequencies of the closed system Hamiltonian $\hat{H}_0$[120], Fig. 2(a). The discussion up to now also holds for systems with *time-reversal symmetry* $\mathcal{T}$, which corresponds to *real symmetric Hamiltonians* (a restricted class of Hermiticity).

*Non-Hermitian scatterers with* $\hat{H}_0 \neq \hat{H}_0^+$. The scenario changes when we consider wave scattering due to scatterers described by non-Hermitian Hamiltonians $\hat{H}_0 \neq \hat{H}_0^+$. The strength of the non-Hermiticity can be controlled by the material gain/loss parameters. Material loss and gain break time-reversal symmetry $\mathcal{T}$, but preserve reciprocity. In the case of lossy ($|d_m(\omega)|^2 < 1$, for each channel) scatterers, the complex zeros move down from the upper half-plane as dissipation increases from zero. At some critical value of dissipation, the zero crosses the real axis at some frequency $\omega_{\text{CPA}}$ leading to perfect absorption of an appropriate coherent incoming waveform, Fig. 2(c). This perfect waveform is determined by the eigenstate of $\hat{S}(\omega_{\text{CPA}})$ corresponding to the zero eigenvalue. This coherent perfect absorber (CPA) is the time-reversed counterpart of a laser for gain media[28,121]. Lasing occurs when a pole touches the real axis at the threshold value of gain, corresponding to an eigenvector of the $\hat{S}$ matrix with an infinite eigenvalue, Fig. 2(c). The interplay between the material gain and loss inside the scatterer can result in *novel wave transport phenomena*. We consider perfect absorption, CPA, VPA, and lasing effects in Section (2).

𝒫𝒯 *-symmetric scatterers.* The evolution of an isolated 𝒫𝒯 - symmetric system is governed by a 𝒫𝒯 -symmetric Hamiltonian, which is symmetrical upon reversal of space 𝒫 and time 𝒯. The 𝒫𝒯 -symmetric optical system can support balanced absorption and amplification, hence respond as a lossless/gainless scatterer, as long as the relative permittivity of the medium satisfies $\varepsilon(\mathbf{r}) = \varepsilon^*(-\mathbf{r})$ [122]. 𝒫𝒯 -symmetric Hamiltonians support a phase transition in the eigenvalue spectrum from entirely real to (partially or completely) complex ones, which occurs at the so-called *exceptional points* (EPs)[123–132]. EPs are specific to non-Hermitian systems and emerge when two (or more) eigenvalues and their corresponding eigenstates coalesce so that the Hamiltonian becomes defective. In the interesting subclass associated with wave scattering due to 𝒫𝒯 -symmetric scatterers, like in the Hermitian case, poles and zeros also occur in complex conjugate pairs, as seen from the interchange between the secular equation for the poles and zeros after performing the combined 𝒫𝒯 -symmetry operation. Nevertheless, as the strength of non-Hermiticity varies, pairs of pole and zero move symmetrically in the complex frequency plane, Fig. 2(d). Eventually, a pole and a zero of the $\hat{S}$ matrix may coalesce on the real axis, and a *CPA-laser*[133–135] regime emerges. In this regime, the system behaves simultaneously as a laser oscillator and a CPA. We consider scattering in 𝒫𝒯 -symmetric systems in Section (3).

**1.7 Quantum noise and the S-matrix approach**

The properties of the S-matrix in the case of 𝒫𝒯 -symmetric systems require special handling. It is known that active systems typically deal with quantum noise associated with enhanced spontaneous emission processes. Noise can be taken into account based on the so-called Langevin forces obeying the fluctuation-dissipation theorem [136]. In the presence of quantum noise, the input and output waves are treated as bosonic operators ($\hat{s}^+$ and $\hat{s}^-$), with the canonical bosonic commutation relations, $[\hat{s}_n^+(\omega), \hat{s}_m^+(\omega')] = \delta_{nm}\delta(\omega-\omega')$, $[\hat{s}_n^-(\omega), \hat{s}_m^-(\omega')] = \delta_{nm}\delta(\omega-\omega')$, and the input-output equation Eq. (6) generalizes to [137]

$$\hat{\mathbf{s}}^- = \hat{S}\hat{\mathbf{s}}^+ + \hat{L}\hat{\mathbf{b}}, \tag{20}$$

where, $\hat{L}$ is the *Langevin operator*, and the operator $\hat{\mathbf{b}}$ represents quantum fluctuations in the system. This approach has been extensively used to describe absorbing and amplifying systems [136,138]. The fluctuation-dissipation theorem yields $\hat{L}\hat{L}^+ = \hat{S}\hat{S}^+ - 1$ [138,139].

Based on this general theory, in the next Sections, we analyze various scattering anomalies corresponding to some of the extreme scenarios captured by the scattering matrix poles and zeros.

## 2. Scattering by extended objects

We will begin our description of scattering anomalies looking into structures extended in one or two dimensions, such as slabs, multilayered structures, and waveguides. Scattering of an incident plane wave or a waveguide mode can be usually reduced to only one or two channels (or four channels for oblique incidence), which makes the problem simpler. In this case, scattering can be characterized by reflection and transmission coefficients, which capture all possible scattering anomalies exhibited by a system.

## 2.1. Perfect absorption (PA)

We have discussed above how the scattering behavior of any linear structure is entirely determined by two basic kinds of special points in the complex frequency plane, zeros and poles of the scattering matrix or its eigenvalues. As we established above, the first type corresponds to perfectly absorbing solutions, while the latter corresponds to QNMs. In a reciprocal Hermitian system, the poles and zeros of $\hat{S}$ always exist in pairs, and their positions are related *via* complex conjugation. Specifically, poles are located in the lower plane [we use the time convention $\exp(-i\omega t)$], while zeros are located in the upper plane, which corresponds to exponential attenuation and divergence in time, respectively.

The presence of losses or gain breaks time-reversal symmetry $\mathcal{T}$ (but preserves reciprocity). An appropriate combination of geometrical design and a specific amount of material loss can push one of the zeros to the real axis, enabling a perfectly absorbing mode at a real frequency, i.e., the possibility of totally absorbing the impinging wave without any scattering for monochromatic excitation. Vice versa, adding gain to the system can push poles towards the real axis until the trajectory of a certain pole intersects it at the *lasing threshold*, giving rise to self-sustained oscillations of the electromagnetic field without incident field at the same frequency[140,141]. Note that the presence of a pole in the upper plane turns the system in an *unstable regime* with the corresponding QNM divergent in time[142].

In most perfect absorbers, the incident energy is delivered *via* a single channel, e.g., by a plane wave incident on one side of the absorber. This is the case of one-port perfect absorbers, such as *Salisbury* and *Dallenbach screens*[143,144]. More recently, diffraction gratings and metasurfaces were shown to enable perfect absorption with a one-side incidence[145–151]. From the fundamental perspective, perfect absorption in all such devices occurs through destructive self-interference, Fig. 3(a)[152–154]. To enable destructive interference, an absorbing material should be placed above a mirror or covered by an anti-reflective coating[155]. This requirement of self-interference imposes the so-called *Rozanov limit* on the absorber thickness[156].

Without self-interference, perfect absorption in one-port structures usually cannot be achieved. Consider, for example, a p- (s-) polarized plane wave incident from vacuum at an angle $\theta$ on a half-space of isotropic material with permittivity $\varepsilon$. The amplitude of the reflected wave in this case reads

$$r_p = \frac{\cos\theta - \frac{1}{\varepsilon}\sqrt{\varepsilon - \sin^2\theta}}{\cos\theta + \frac{1}{\varepsilon}\sqrt{\varepsilon - \sin^2\theta}}, \tag{21a}$$

$$r_s = \frac{\cos\theta - \sqrt{\varepsilon - \sin^2\theta}}{\cos\theta + \sqrt{\varepsilon - \sin^2\theta}}. \tag{21b}$$

In the lossless scenario, at the *Brewster angle,* the reflection of a *p*-polarized plane wave vanishes. The presence of loss forbids perfect transmittance due to the impedance mismatch between air and the lossy medium, whose normalized impedance $Z = \frac{1}{\varepsilon}\sqrt{\varepsilon - \sin^2\theta_i}$ has a non-vanishing imaginary part associated with the electric loss in the dielectric, and making perfect absorption impossible.

Thus, achieving one-sided perfect absorption without backing mirrors is challenging. This requires excitation of both electric and magnetic currents within an absorbing conductive sheet, creating destructive interference similar to a Huygens source[153]. There are non-trivial examples of one-port systems that allow perfect absorption to occur without self-interference. One such example is a half-space made of an anisotropic medium with the optical axis being normal to the surface. In that case, even in the presence of losses, the reflection of a *p*-polarized plane wave can become zero, giving rise to complete absorption of the incident wave inside the lossy material. Such interference-free total absorption was demonstrated in mid-IR for hexagonal boron nitride[157] and theoretically predicted in asymmetric hyperbolic media[158]. Photonic structures lacking $C_2^z$ symmetry, i.e., 180-degree rotation around the out-of-plane axis, were also shown to exhibit nearly perfect one-sided absorption[159].

## 2.2. Coherent perfect absorption (CPA)

In the above examples of structures exhibiting zeros of S-matrix eigenvalues (perfect absorption), incident energy was assumed to enter a system *via* one specific channel. That situation corresponds to the most practical case of an absorber illuminated, e.g., by the Sun or a lab light source, from one side. Coherent perfect absorber (CPA) concept generalizes this scenario to an arbitrary incoming waveform $\mathbf{s}^+$, consisting of two or more waves, such as waves incident on opposite faces of an open slab or film[28,121,160,161]. To achieve perfect absorption, one must choose appropriate values for system parameters and the operating wavelength—as well as the suitable input waveform $\mathbf{s}^+$, including the intensities and relative phases of the input beams. Conversely, by adjusting the relative phase of the inputs, one can switch to the regime of suppressed absorption, caused by the *constructive* interference of exiting waves[28,161,162]. Interestingly, this regime corresponds to the time-reversed operation of a laser (anti-laser) [28,163–166]. We caution, however, against a direct literal understanding of this relationship, because it does not take into account nonlinear processes and spontaneous emission, which are essential in the laser operation. Note also that the CPA can be considered as the generalization of the concept of critical coupling [33,167].

The mathematical condition for coherent perfect absorption is that output field amplitudes vanish for a certain non-zero set of input amplitudes $\hat{S}\mathbf{s}^+_{\text{CPA}} = 0$. For this to happen, at least one of the eigenvalues of $\hat{S}$ must be zero, just like in the case of a single-port absorber. The difference lies in the structure of the "CPA eigenmode" eigenvector $\mathbf{s}_{\text{CPA}}$, which can have several non-zero components meaning that the system should be illuminated in more than one channel to achieve coherent perfect absorption, Fig. 3(c).

To illustrate the phenomenon of coherent perfect absorption, we consider the simplest case of a two-port planar structure illuminated from the two sides. Ignoring the polarization degree of freedom, the inputs and outputs are related *via* (for normal incidence)

$$\begin{pmatrix} s_1^- \\ s_2^- \end{pmatrix} = \hat{S} \begin{pmatrix} s_1^+ \\ s_2^+ \end{pmatrix}, \quad \hat{S} = \begin{pmatrix} r_{11} & t_{12} \\ t_{21} & r_{22} \end{pmatrix}. \tag{22}$$

Here $s_i^+$ and $s_i^-$ respectively denote the input and output wave amplitudes in the i-th channel; $r_{ii}$ are the reflection coefficients in each port, and $t_{ij}$ are the transmission coefficients. If $\hat{S}$ obeys reciprocity[101], then $t_{12} = t_{21} = t$. If, also, the two-port structure is symmetric under mirror reflection that exchanges the two ports, then $r_{11} = r_{22} = r$ and the two eigenvalues of $\hat{S}$ are then $r \pm t$. The corresponding eigenvectors are $\mathbf{s}_{\text{CPA}}^+ \propto (1, \pm 1)$, representing *symmetric and antisymmetric inputs* of equal intensity. Now, if $r \pm t = 0$ for a planar structure at some frequency, then illumination of it with the corresponding CPA eigenmode will lead to complete absorption.

The absorption in a CPA turns out to be highly sensitive to variations in the conditions of illumination, in particular in the relative phase between the components of the exciting field. To characterize a CPA, it is convenient to define the *joint absorption coefficient*[162]

$$\mathcal{A} \equiv 1 - \frac{|s_1^-|^2 + |s_2^-|^2}{|s_1^+|^2 + |s_2^+|^2}. \tag{23}$$

Here, $|s_1^+|^2 + |s_2^+|^2$ is proportional to the total input intensity and $|s_1^-|^2 + |s_2^-|^2$ to the total output intensity; thus a CPA has $\mathcal{A} = 1$. Fig. 3(c, right panel) features a typical plot of $1-\mathcal{A}$ for the symmetric (blue curve) and anti-symmetric (orange curve) modes of a uniform dielectric slab near a CPA frequency. CPA occurs whenever one of the curves falls to zero. An intriguing feature of this plot is that at the CPA frequency, while one of the incident modes yields total absorption, the other mode strongly scatters due to constructive interference outside the system, thus suppressing the absorption.

Zeros of different eigenvalues of $\hat{S}$ typically occur at distinct frequencies, meaning that if a certain incident waveform $\mathbf{s}_{\text{CPA}}$ experiences coherent absorption, the orthogonal waveform does not. However, there are examples of two-port mirror-symmetric structures with doubly degenerate CPA mode[168,169], in which both eigenvalues of $\hat{S}$ can be zero at a certain frequency (double zero). Note that the coalescence of two (or more) poles leads generically to an exceptional point (EP) [122,126,170], as discussed in Section (3). The absorption becomes insensitive to the relative phase and amplitudes of the two input waves so that even a single input wave incident on either port would be perfectly absorbed. Such a system presents an example of one-sided perfect absorption without backing mirrors.

CPAs have been studied and realized in a wide variety of geometries, including planar structures (slabs, diffraction gratings, metasurfaces, and thin films)[28,160,161,171] and guided-mode structures[172–175] compatible with integrated nanophotonics applications. This approach has been utilized for controlling scattering and radiation from optical nanostructures and nanoantennas [175,176], and also for boosting wireless power transfer between two antennas [177]. Different techniques have been devised for increasing or decreasing the absorption bandwidth[178–180], and for achieving perfect absorption under strongly-asymmetric beam intensities[181]. Also, CPA has been extended to systems exhibiting strong light-matter interaction[182]. A recent study demonstrates CPA in a disordered medium implementing a random anti-lasing regime [164].

The operation of CPAs in the quantum optical case, when a few-photon state represents an incident field, has been also studied[183–185]. For example, in Ref.[185] absorption of two-photon states with 40% efficiency has been demonstrated. The framework of the electromagnetic $\hat{S}$ matrix does not accurately capture this situation, since a lossy structure (beam splitter) may induce transitions between few-photon and single-photon states, which are unmonitored by the electromagnetic $\hat{S}$ matrix.

The phenomenon of coherent perfect absorption may be useful in a variety of applications, ranging from all-optical data processing to enhanced photocurrent generation. CPAs can be highly sensitive to the parameters of the input waves, enabling attractive opportunities for the flexible control of light scattering and absorption[161,186,187]. In a two-port CPA, for instance, one input can be regarded as a signal beam, and the other as a control beam. A coherent control beam serves as a resource enabling efficient control of optical absorption, even in the linear regime. This opens the door to novel low-energy logical circuits and small signal amplifiers[186,187] without the use of non-linear materials that rely on high intensity of incident radiation to achieve a nonlinear response. Moreover, coherent perfect absorption can be employed for the processing of binary images and pattern recognition[188]. Coherent absorption effects may prove very useful for improving photocurrent generation in photoelectrochemical systems[189]. Finally, the ability to control and remotely trigger coherent perfect absorption of single- or few-photon states[183,184] may be promising for quantum information processing.

Recently, these studies on coherent all-optical tuning and manipulation have inspired research efforts studying the possibility of boosting the performance of antenna systems. It has been shown that the same principle that underlies the operation of CPAs can be employed to improve the efficiency of antenna energy transfer, and as a result enhance the efficiency of wireless power transfer (WPT) systems [190]. More specifically, by coherently exciting the receiving antenna with an auxiliary signal, tuned in sync with the impinging signal from the transmitting antenna, it is possible to largely enhance the robustness and overall efficiency of the system. This additional signal improves energy transfer through constructive interference with the impinging wave, compensating any imbalance in the antenna coupling, without having to modify the load. In another work, it has been explored that coherent excitation of an antenna by two dipole sources allows *controlling the multipoles* resulting in boosted superradiance and enhanced directivity [176]. Interestingly, it makes possible to excite nonradiative field configurations, and turning them on/off on demand. Remarkably, this approach also allows reducing the quenching effect and enhancing the radiation efficiency without changing the antenna's geometry via constructive or destructive interference of the excited QNMs.

## 2.3. Virtual perfect absorption (VPA) and virtual gain

In the previous section, we have discussed the perfect absorption effect for harmonically oscillating incident waves, which can occur when an S-matrix zero is located at a real frequency $\omega$. However, in the case of lossless objects obviously, the S-matrix zeros [unlike scattering zeros, see Section (1.4)] cannot arise at any real frequency (except for bound states in the continuum, where the zeros are canceled out by the complex conjugate pole) and are restricted to complex frequencies. These frequencies correspond to field solutions $\mathbf{E}_{zero}(r,t)$ with time dependence described by an exponentially growing amplitude $e^{\omega^r t}$, such that

$$\hat{S}\mathbf{E}_{zero} = 0 \tag{24}$$

In light of this fact, the following question arises: can we somehow engage these complex modes to mimic perfect absorption in lossless systems?

It turns out that, instead of adding loss to the system to push a complex zero towards the real axis, we can tailor the incident field in time, such that its temporal profile matches the exponentially diverging mode associated with a complex zero over a finite interval of time, giving rise to *virtual perfect absorption*[191]. During this transient, the scattering from the structure totally vanishes, as if the structure were perfectly absorbing, despite the absence of material loss, Fig. 3(d). Instead, the energy is efficiently stored inside the open cavity without scattering. When the exponentially growing incident field is stopped, it gives rise to the release of the energy stored in the system. This behavior can be interpreted as transient destructive interference of the scattered waves with the waves that would leak from the cavity.

The effect of virtual absorption can be reproduced in any lossless electromagnetic structure, provided that the spectral position of its zero is known, and the corresponding spatial and temporal profile of the incident field can be created. It is also robust to inevitable material dissipation and material dispersion because the position of poles and zeros entails information about this dispersion. Finally, this effect is not limited to optics. Indeed, the scattering matrix formalism is equally applicable to acoustics[192] and elastodynamics and single-particle quantum mechanical problems. Since the Hamiltonians of quantum mechanical systems are naturally Hermitian, such systems have zeros of S-matrix in the upper half-plane of the complex energies. Therefore, properly shaping the temporal wavefunction of an incident particle[193] is expected to enable efficient storing of the wavefunction inside a potential without scattering. This observation suggests that this effect can be employed for efficient low-energy light storage and release on demand. The idea of coherent virtual absorption has been realized experimentally with elastodynamic waves [192].

Finally, if in the transient the incoming signal exponentially decays faster than the cavity decay rate, the output signal [along with the energy and momentum flux] can be *larger than the input one*. In this case, the stored energy becomes effectively "negative", as one would have in a system with real gain. This effect, which we refer to as "virtual gain", has been recently utilized for tailoring the optical *pulling* force in paraxial fields where otherwise the radiation force is always positive due to the momentum conservation law. Somewhat related, a complex excitation with negative imaginary frequency has been also suggested to improve flat imaging devices [194].

## 2.4. Unconventional lasing scenarios

Non-Hermitian photonic structures involving optical gain and loss, often result in anomalous scattering behaviors that are absent in passive systems[122,127,195–199]. Given the ubiquity of gain and loss processes in various electromagnetic structures and devices, such as lasers, absorbers, and isolators, systematic studying their optical response *via* the scattering poles and zeros is highly desirable.

Poles of the scattering matrix hitting the real-frequency axis correspond to lasing[141,200–203]. From the classical electromagnetics perspective, *lasing can be viewed as the time-reversed process of perfect absorption*, i.e., emission of radiation from an otherwise non-illuminated structure[28,140]. This picture does not account for optical or electrical pumping of the gain medium that provides population inversion and amplification of light at the emission frequency. Instead, fixed population inversion is simply assumed in this picture, which is reflected in the negative imaginary part of the gain medium permittivity [203–208]. For a rigorous analysis of a laser, this approach is not sufficient, and the rate equations [209–211] or Maxwell-Bloch equations [208,212] can be used.

An important difference between lasing and perfect absorption in terms of the poles and zeros analysis is that, when a zero moves below the real frequency axis as more absorption is added to the system, the perfectly absorbing response weakens, i.e., absorption is maximized for a specific value of loss that balances radiation loss at the critical coupling condition. In contrast, adding gain to a lasing system implies in most cases a further increase in the amplitude of its emission. In a more rigorous description of lasers based on Maxwell-Bloch equations[208,212], the population inversion of the gain medium itself depends on the electric fields in the system, which makes the problem nonlinear[213,214]. For this reason, linear description of a system with gain is only applicable below the lasing threshold. Nevertheless, the linear theory still allows one to evaluate whether the system can operate as a laser at all. If at least one scattering pole lies in the upper half of the complex plane (or at the real axis), lasing oscillations arise.

In a typical representation of the laser operation, adding gain into the system shifts the poles towards the real axis. Adding loss, on the contrary, moves poles away from the real axis making lasing harder to reach. Such a naive picture originating from the single-mode laser rate equations, however, turns out to be not always correct. In particular, a non-uniform increase of gain may lead to the death of laser generation[215,216]. The first discovery of this counter-intuitive *gain-induced laser turn-off* was reported in[215], where the authors theoretically studied lasing from two coupled slabs containing gain medium pumped non-uniformly. First, one cavity of the system was pumped until it started lasing. Then, the second cavity was pumped while the pumping (and therefore gain) of the first one kept unchanged. It turned out that, upon pumping of the second cavity, the output laser emission began to drop until the laser eventually turned off, Fig. 4(a). Such unusual behavior is induced by the presence of *exceptional points* (EPs)[217], at which two scattering poles of the systems coalesce. We stress that this effect, although being related to the non-linear threshold dynamics of a laser, can be totally captured by linear equations. Laser turn-off was experimentally demonstrated in a configuration involving a pair of coupled microdisk lasers[218] and in a pair of active RLC-circuits[142].

Not only the increase of gain can lead to laser turn-off, but, vice versa, *increase of loss can lead to the onset of lasing*. Similarly to gain-induced laser turn-off, this effect may be caused by EPs in the spectra of systems with the non-uniform loss[219]. Such behavior was also predicted for homogeneous structures without EPs, but with dispersive absorbing medium[220]. An increase of the loss parameter in this case not only increases the rate of energy dissipation but also changes the effective refractive index of the cavity, potentially pulling the scattering pole into the region of the gain medium and enabling lasing.

In general, for a real-valued EP (two poles coalesce at the real axis), the variation of either gain or loss (or both of them) can cause EP breaking with the formation of two QNMs, lossy and "gainy" (i.e., located in the upper plane). An analysis involving nonlinearities (e.g., a saturable gain and a saturable loss) shows that imbalances between the saturation intensities cause a *bifurcation* in the vicinity of the exceptional point [221]. Such bistable lasers are interesting for possible applications as optical "flip-flop" memory devices [222–224].

For our example of a slab uniformly filled with a gain medium, the eigenvectors of the lasing modes have the form $\mathbf{a}_{las} = (1, \pm 1)$ corresponding to symmetric and anti-symmetric eigenmodes. Such modes produce the equal distribution of emission intensity leaving the system from the left and right sides. In fact, any reciprocal slab without ideal mirrors (such that $t \neq 0$) produces non-zero emission at both sides at the lasing threshold. However, adding magneto-optical material to the design of a mirrorless multilayered structure has been shown to enable *unidirectional lasing*, Fig. 4(b)[225]. The coexistence of *non-reciprocity* and slow light modes leads to giant asymmetry of transmission, resulting in unidirectional mirrorless lasing not available in reciprocal active cavities. Another approach towards unidirectional singularities based on PT-symmetry was proposed in[226,227], as discussed in more detail in the next section.

To reduce radiation losses and hence reduce lasing threshold one can engineer the structure supporting resonant dark modes [228,229]. For example, a low lasing threshold has been demonstrated for collective resonances in 2D plasmonics arrays [230,231]. In an extreme scenario, we can realize bound states in the continuum (BICs), which are QNMs uncoupled from the continuum of radiation modes [118,232–234]. BICs have no radiation losses, and hence the Q-factor of a structure without no material loss can be unlimitedly large. Therefore, such structures are promising for low-threshold lasers [233,235]. BICs are discussed in more detail in Section 4.

Finally, we note that this criterion of a laser regime, i.e. the presence of a pole (of scattering amplitude or S-matrix) at the real frequency axis corresponds to the so-called classical condition[33,140]. For nanolasers, or lasers having dimensions less than the radiation wavelength, there is the *quantum threshold condition*[236,237], which requires the number of stimulated photons in the mode to overcome the number of spontaneous ones[236]. The spontaneous emission also causes the spectral broadening of the laser regime [139], so that the spectral width is not $\delta$-like but rather broadened.

## 3. Exceptional points

Generally speaking, EPs arise when two or more eigenvalues along with their eigenvectors of a linear operator coalesce. This leads to reducing the size of the space spanned by the eigenbasis, and the operator becomes defective [197]. Such exceptional points can appear in a wide class of non-Hermitian operators. Physically, EPs correspond to coalescence of poles of the S-matrix (or scattering amplitudes), which can happen either in *the complex plane* below the real axis (in passive systems) or *at real frequencies* (in active systems). We also note that recently reported studies demonstrate an intriguing way to EPs-governed scattering based on a coalescence of S-matrix *zeros* [238], exploiting the physics of the coherent perfect absorption (CPA) regime. EPs find unprecedented capabilities for sensing applications [132] and tailoring of unconventional lasers [239–242]. Also, active systems supporting EPs typically become nonlinear above the laser threshold, which makes them non-reciprocal in the breaking phase [243] and giving rise to several possible applications. Below in this section, we focus our analysis at the EPs in $\mathcal{PT}$-symmetric active structures.

### 3.1. $\mathcal{PT}$ -symmetric structures

In recent years, there has been a great interest in the so-called $\mathcal{PT}$-symmetric structures, a particular class of non-Hermitian systems that exhibit balanced regions of gain and loss[122,124–127,244–250]. The concept of $\mathcal{PT}$-symmetry originated in the context of theoretical quantum physics[251,252], after the pioneering work of Bender and Boettcher, showing that a large class of non-Hermitian Hamiltonians can exhibit entirely real spectra as long as they respect $\mathcal{PT}$-symmetry. In general, a Hamiltonian is said to be $\mathcal{PT}$-symmetric if it is invariant under simultaneous parity- ($\mathcal{P}$) and time- ($\mathcal{T}$) reversal operations[126,250–252]. Almost a decade later, it was realized that the concept of $\mathcal{PT}$-symmetry can be extended to the realm of optics[123,124,247], where time-reversal involves mapping optical gain to loss. A non-magnetic optical structure is $\mathcal{PT}$-symmetric if the dielectric function satisfies

$$\varepsilon(\mathbf{r}) = \varepsilon^*(-\mathbf{r}). \tag{25}$$

Despite being non-Hermitian, $\mathcal{PT}$-symmetric systems exhibit certain features similar to Hermitian systems. Most importantly, $\mathcal{PT}$-symmetric Hamiltonians can have purely real eigenvalues within a parameter domain called the exact $\mathcal{PT}$ phase regime.

For certain parameter ranges, the system undergoes a *phase transition*, and its eigenvalue spectrum becomes complex. In the latter scenario, $\mathcal{PT}$-symmetry is spontaneously broken given that the eigenstates are no longer invariant under simultaneous parity and time-reversal even though the governing Hamiltonian commutes with the $\mathcal{PT}$ operator[124,125,251,252]. The $\mathcal{PT}$ symmetry phase transition in the parameter map of the system occurs at the *exceptional point* (EP)[123–132] at which, not only the eigenvalues but also their associated eigenvectors coalesce.

Perhaps, the simplest example of a $\mathcal{PT}$-symmetric structure is a coupler composed of two evanescently coupled guided-wave elements, i.e., optical waveguides or cavities, which are identical in any sense except for the presence of gain in one element while the other element exhibits the same amount of loss, see Fig. 5(a). Figures 5(b,c) depict the real and imaginary parts of the Hermitian eigenvalues of such a system, which clearly show a phase transition at the exceptional point singularity. Quite interestingly, in direct analogy with Hamiltonian systems, $\mathcal{PT}$-symmetric scattering settings can undergo a phase transition that is associated with the bifurcation of the S-matrix eigenvalues from the unit circle in the complex domain. This phenomenon is explained in greater detail in the following.

For a general $\mathcal{PT}$-symmetric scattering system, the condition of Eq. (25) implies that the S-matrix at an arbitrary frequency $\omega$ satisfies[253]

$$(\mathcal{PT})\hat{S}(\omega^*)(\mathcal{PT}) = \hat{S}^{-1}(\omega), \tag{26}$$

where, $\mathcal{P}$ and $\mathcal{T}$ are representations of the parity and time-reversal operations[134]. Here, the parity operator $\mathcal{P}$, is a linear operator subject to $\mathcal{P}^2 = \hat{I}$, and the time-reversal operator $\mathcal{T}$, is an anti-linear operator, which represents the complex conjugation[254]. It is worth noting that in a Hermitian system, the scattering matrix satisfies $\mathcal{T}\hat{S}(\omega^*)\mathcal{T} = \hat{S}^{-1}(\omega)$ which is associated with the important result of unimodular S-matrix eigenvalues for passive lossless systems when considering real frequencies. In the case of a $\mathcal{PT}$-symmetric system, one can show that the S-matrix eigenvectors ($\hat{S}\upsilon_n = \sigma_n \upsilon_n$) are governed by $\hat{S}(\omega^*)(\mathcal{PT}\upsilon_n) = \frac{1}{\sigma_n^*}(\mathcal{PT}\upsilon_n)$. Assuming a one-dimensional system, in the exact $\mathcal{PT}$ phase regime, the two eigenvectors respect $\mathcal{PT}$ symmetry ($\mathcal{PT}\upsilon_\pm = \upsilon_\pm$) thus the eigenvalues are unimodular at real frequencies, i.e., $|\sigma_\pm|^2 = 1$, as in Hermitian systems. In the broken $\mathcal{PT}$ phase regime, on the other hand, the $\mathcal{PT}$ operator exchanges the two eigenvectors, i.e., $\mathcal{PT}\upsilon_{1,2} = \upsilon_{2,1}$, which instead results in $\sigma_1 \cdot \sigma_2^* = 1$. In this latter scenario, the two eigenvalues are no longer unimodular as amplifying and dissipating eigenstates arise.

Next, we consider a one-dimensional bilayer structure of length $L$ composed of two slabs of gain and loss materials with the complex refractive indices of $n_{1,2} = n \pm ig$, Fig. 5(d). In general, the S-matrix eigenvalues of this system can be written in terms of the reflection ($r_{11}$, $r_{22}$) and transmission ($t$) coefficients as follows:

$$\sigma_{\pm} = \frac{r_{11}+r_{22}}{2} \pm \sqrt{t^2 + \left(\frac{r_{11}+r_{22}}{2}\right)^2}. \quad (27)$$

The eigenvectors of the system, without a particular choice of normalization, can be defined as $\upsilon_{\pm} = (1, \rho_{\pm})^T$, where, the ratio of the eigenvectors $\rho_{\pm}$ is:

$$\rho_{\pm} = \frac{r_{22}-r_{11}}{2t} \pm \sqrt{1 + \left(\frac{r_{22}-r_{11}}{2t}\right)^2}. \quad (28)$$

Figure 5(e) depicts the S-matrix eigenvalues of the $\mathcal{PT}$-symmetric slab when the frequency is scanned on the real axis. As this figure clearly indicates, at a critical frequency the eigenvalues bifurcate from the perimeter of the unit circle; one moving inside while the other one traveling outside the unit circle. Also, for one of the eigenvectors the ratio $\rho$ approaches zero, while for the other one, this ratio grows indefinitely. The bifurcation of the eigenvalues manifests itself in the scattering coefficients of the system. In fact, below the phase transition point, the scattering parameters involve two resonance features associated with the even and odd modes of the $\mathcal{PT}$ Fabry-Perot. By passing the phase transition point, however, the two resonances merge.

Following the theme of the previous sections, it is of interest to find the general trend of the poles and zeros of the S-matrix eigenvalues of the $\mathcal{PT}$-symmetric Fabry-Perot. As opposed to a single slab of gain or loss material, where any change in the gain-loss simply up- or down-shifts the poles and zeros in the complex frequency plane, here the effect is much more prominent. In fact, by increasing the gain-loss parameter $g$, the zeros of the adjacent even and odd eigenstates approach each other along the real frequency axis until meeting at an exceptional point, and afterward they repel each other along the imaginary frequency axis. Given that the poles appear as complex conjugate zeros, the poles are governed by a similar trend. This behavior is illustrated in Fig. 5(f), for a two-port bilayer gain/loss structure. An interesting scenario occurs at a critical point where a pole-zero pair meet at the real frequency axis, giving rise to simultaneous lasing and coherent perfect absorption (CPA-laser). Notably, this occurs even though there is zero *net* gain or loss in the structure.

The scattering behavior of $\mathcal{PT}$-symmetric structures becomes more complex in higher dimensions. A basic example of such geometries is a Janus particle with balanced regions of gain and loss[244] or a dimmer of two coupled dielectric spheres with equal gain and loss[255,256]. Interestingly, it has been shown that $\mathcal{PT}$-symmetric particle dimmers exhibit highly anisotropic scattering patterns depending on the angle of incidence[255]. Also, for a light beam impinging the structure along the interface of the gain and loss regions, the scattering pattern is deflected even though both regions have equal indices of refraction[244]. Further studies have theoretically demonstrated unidirectional cloaking based on $\mathcal{PT}$-symmetric material cloaks[257,258]. These interesting properties occurring in basic $\mathcal{PT}$-symmetric structures, reveal a potential for utilizing the interplay of gain/loss with refractive index in complex geometries in order to manipulate the propagation of light.

We now focus on a dimer of coupled dielectric spherical particles with equal gain and loss. Assuming that the two particles are much smaller than the wavelength, this problem can be treated through the dipole approximation[259,260]. In this approximation, the electric dipole moment of each particle, say particle $i = 1,2$, is related to the total electric field at the location of the particle $\mathbf{r}_i$ through $\mathbf{p}_i = \alpha_i \mathbf{E}^{tot}(\mathbf{r}_i)$ where $\alpha_i$ represents the electric polarizability of the particle. On the other hand, the total electric field at the location of each particle can be considered as a sum of the incident field as well as the contribution from the other particle, thus $\mathbf{E}^{tot}(\mathbf{r}_i) = \mathbf{E}^{inc}(\mathbf{r}_i) + \mathbf{G}(\mathbf{r}_i, \mathbf{r}_j) \cdot \mathbf{p}_j$ where $\hat{\mathbf{G}}(\mathbf{r}_i, \mathbf{r}_j)$ is the dyadic Green's function. Therefore, one can solve for the dipole moments of the two particles in terms of the incident field. Assuming that the incident field is equal at the location of the two particles, one can consider effective polarizability for the $\mathcal{PT}$ dimmer $\alpha_{tot} = (p_1 + p_2)/E^{inc}$, which is found to be[261]:

$$\alpha_{tot} = \frac{\alpha_1^{-1} + \alpha_2^{-1} + 2G}{\alpha_1^{-1}\alpha_2^{-1} - G^2}. \tag{29}$$

Here, we consider only a transverse or longitudinal polarization for the incoming wave, where its electric field component is orthogonal or parallel to the particle separation, respectively. The polarization information is hidden in the scalar Green's function $G$.

Let us consider the particles of radii $a$, with separation $d$, and of complex conjugate permittivities $\varepsilon_{1,2} = \varepsilon_r \pm i\varepsilon_i$ as shown in Fig. 5(g). In the long-wavelength limit, and for $a \ll d$, the zeros and poles of the total polarizability can be obtained analytically to lowest orders in $a/d$. Interestingly, it is found that the poles and zeros form two tangential circles in the complex permittivity plane, i.e., $(\varepsilon_r, \varepsilon_i)$ [261]. For a longitudinal excitation, the two circles are outer tangent, while for the transverse excitation they are inner tangent [Fig. 5(h,i)]. In both scenarios, the pole-zero cancellation point happens near the plasmonic resonance, i.e., $\varepsilon_r = -2$ and reveals the existence of a bound state in the continuum (BIC)[261,262]. The behavior of $\mathcal{PT}$-symmetric dielectric particles has also been investigated away from the plasmonic resonance[256,263]. It has been shown that similar to the 1D case, here the scattering matrix undergoes a phase transition and becomes non-unitary at a critical gain/loss contrast[264].

## 3.2. Laser-Absorber

The simultaneous appearance of laser and absorber modes was first predicted in a one-dimensional $\mathcal{PT}$-symmetric structure[134,135], while it has also been reported in $\mathcal{PT}$-symmetric microring resonators in Ref. [133]. Experimental demonstrations of $\mathcal{PT}$-symmetric laser-absorbers, consisting of straight waveguides on substrates, were recently reported in two independent papers[163,265]. In Ref.[266], a halving of the emission peaks was observed, indicating the occurrence of the $\mathcal{PT}$-broken phase, though there was no direct evidence that the system was operating as a CPA-laser. In Ref.[163], however, the input and output intensities were directly measured, and for different phase offsets of the inputs, the authors observed both a maximum and a minimum of the relative output intensity, with a substantial contrast of around 30 dB. As nonlinear laser dynamics arising above the laser threshold may significantly alter the $\mathcal{PT}$ symmetry, the pump intensity was maintained near the lasing threshold, Fig. 6(a).

## 3.3. Unidirectional invisibility

$\mathcal{PT}$-symmetric structures exhibit strongly asymmetric reflection when operating near the $\mathcal{PT}$-breaking threshold[267,268]. This behavior can be exploited for asymmetric interferometric light-light switching[181] and one-sided perfect absorption[269] [the structure presented in Fig. 6(b)]. In Ref.[181], a $\mathcal{PT}$-symmetric metawaveguide [similar to one depicted in Fig. 6(b)] was realized close to the exceptional point, where strong reflection asymmetry near the exceptional point allowed for a CPA eigenmode with strongly asymmetric incidence, allowing a strong signal beam to be manipulated by a much weaker control beam. An extinction ratio of up to 60 dB was observed, with a 3:1 intensity ratio between signal and control beam.

An interesting example of a one-dimensional $\mathcal{PT}$-symmetric structure is a periodic arrangement of gain and loss slabs that exhibits interesting properties such as *unidirectional invisibility*[267,268,270]. Such an arrangement can be obtained by extending a standard bilayer grating, which involves a periodic stack of two slabs with refractive indices $n_1$ and $n_2$, to a four-layers grating with complex refractive indices $n_1 + ig$, $n_1 - ig$, $n_2 + ig$, $n_2 - ig$. Clearly, this geometry respects the necessary condition of $\mathcal{PT}$ symmetry when the reference axis of symmetry is properly chosen. Alternatively, one can consider a sinusoidal modulation of the refractive index and gain-loss profiles in a constant background, defined as $n(z) = n_0 + n_R \cos(2\pi z / \Lambda_0) + i n_I \sin(2\pi z / \Lambda_0)$, where $\Lambda_0$ represents the modulation period. This model is analytically solvable for weak modulations $(n_R, n_I \ll n_0)$ and when operating near the Bragg wavelength $\lambda_B \approx 2\Lambda_0$ [271]. Under such conditions, the evolution of the slowly varying envelopes of plane waves propagating towards the forward and backward directions can be described with coupled-mode equations[272]. Interestingly, the coupling between the forward and backward waves occurs at different rates of $\frac{k_0}{2}(n_R \pm n_I)$ [271]. This can be compared with a standard Hermitian grating where the counter-propagating waves are coupled at an equal rate of $\frac{k_0}{2} n_R$ [272]. Surprisingly, here, for the critical choice of $n_R = n_I$, the backward propagating wave becomes completely decoupled from the forward while the coupling in the reverse processes occurs at twice the rate of an ordinary grating. As a result, the $\mathcal{PT}$ grating does not leave any footprint on a wave propagating towards one direction while it can be highly reflective in the opposite direction. For a finite $\mathcal{PT}$ grating of length $L$, at the symmetry-breaking threshold, i.e., $n_R = n_I$, and for excitation at Bragg frequency, a simple expression can be obtained for the S-matrix[271]:

$$S = \begin{pmatrix} 0 & 1 \\ 1 & ik_0 L n_I \end{pmatrix}, \tag{30}$$

which clearly shows the one-way invisibility of the system.

Unidirectional invisibility was first experimentally observed in time-domain lattices established through successive interference of optical pulses propagating through two coupled fiber loops[246]. In this arrangement, external phase modulators and amplifiers were used to induce a complex potential for light pulses. Unidirectional reflectionless has been also demonstrated in a passive silicon waveguide with a periodic modulation of its dielectric permittivity and loss[267]. In such a setting, segments of Ge/Cr have been incorporated periodically to create a loss modulation, Fig. 6(b). It should be noted, however, that the transmission of such a system is below unity due to the absence of gain. Also, passive $\mathcal{PT}$-symmetric waveguides have been also realized in organic thin-film waveguides, and significant asymmetric reflection has been observed[273].

### 3.4. Single-mode lasing

As mentioned in previous sections, a striking feature of $\mathcal{PT}$-symmetric systems is the emergence of an abrupt phase transition in their eigenstates. Lately, it has been noted that this symmetry breaking threshold varies for different modes of a multimode system[274]. This interesting property is of great importance in lasing systems where the phase transition threshold is associated with the onset of lasing. Considering a multimode laser coupled to a counterpart lossy cavity, in the framework of coupled-mode theory each mode will pair up with a corresponding mode in the lossy partner. In this case, based on the gain-loss contrast experienced by each mode and the level of coupling, two regimes can be distinguished for each pair of modes. Modes in the unbroken $\mathcal{PT}$-symmetry regime live equally in the gain and loss regions and thus remain neutral while for a pair in the broken $\mathcal{PT}$ phase regime, one mode is located mostly in the active cavity while the other mode remains confined to the lossy cavity[274]. As a result, through proper design, one of the broken-symmetry modes can lase while all other modes being prevented from lasing.

The mode selection property of parity-time symmetric systems has been experimentally demonstrated in optically-pumped semiconductor coupled microring lasers[242] [see Fig. 6(c), upper and middle panels]. In such a system, gain and loss have been controlled through selective pumping of the microring resonators. In a different effort, a microring laser with periodic modulation of loss segments has been also demonstrated[241] [see Fig. 6(c), lower panel]. In both works, undesired longitudinal modes of microrings have been filtered out to achieve single-mode lasing. A similar method can be utilized to filter transverse modes of multimode cavities, as demonstrated in large area microring lasers[245] as well as in stripe lasers[275] and microdisk lasers[276].

### 3.5. Passive PT-symmetric systems and exceptional points

As discussed above, the $\mathcal{PT}$-symmetry condition in optics requires the real [$n(\mathbf{r})$] and imaginary [$k(\mathbf{r})$] parts of the refractive index [$\tilde{n}(\mathbf{r}) = n(\mathbf{r}) + ik(\mathbf{r})$] to be even and odd functions of coordinates ($\mathbf{r}$). The most general way to satisfy this condition relies on the presence of gain ($\text{sign}[k(\mathbf{r})] < 0$) compensating the material loss. However, from the engineering perspective, the gain can cause instability issues and noise, which is always attributed to active media[277–281]. This also causes many difficulties in the realization of PT-symmetric quantum optics because the gain becomes random at the few-photons level due to the spontaneous emission[282–284].

Nevertheless, there are two approaches to mimic the PT-symmetry conditions in passive systems. The first approach to passive PT-symmetry ($\text{sign}[k(\mathbf{r})] > 0$) based on *gauge transformation* in a system of two (or more) coupled resonant cavities where one cavity exhibits loss, while the other is lossless[247]. Remarkably, the first demonstration of PT-symmetry effects in optics was relying on this technique[247]. Another approach to passive PT-symmetry relies on passive periodic structures with [$n(\mathbf{r})$] and [$k(\mathbf{r})$] varying with equal amplitude, such as e.g., in the one-dimensional case $\Delta\tilde{n}(r) = \Delta n \cos(qr) + i\Delta k \sin(qr)$ [273,285,286]. EPs in such passive systems arise due to the unbalanced competition between coupling strength and loss contrast of two or more states. However, although the discussed passive PT-symmetry scenarios are promising for light manipulation, they remain passive and are not indeed PT-symmetric. Their relation to PT-symmetry requires postprocessing.

## 4. Embedded eigenstates

Trapping of light in electromagnetically open structures is important for various applications in optics, including all-optical memory, lasing, sensing and so on. QNMs allow light trapping due to the resonance effect. In Section (1.5) we discussed that if a QNM has the complex frequency $\omega = \text{Re}[\omega] + i\text{Im}[\omega]$, it provides light trapping for the QNM lifetime $\tau = -1/2\text{Im}[\omega]$ (since QNM is characterized by negative $\text{Im}[\omega]$) before the energy gets decay radiatively and dissipatively by a factor of ~2.7. During this decay, the QNM oscillates approximately $Q = -\text{Re}[\omega]/2\text{Im}[\omega]$ times in transient after abrupt turning-off of the excitation. In the frequency domain, higher losses, radiative and dissipative, pushes the pole deeper to the lower complex plane leading to the resonance broadening. Hence, ideal light-trapping requires a system to be Hermitian with canceled out radiation loss, which can be achieved via the VPA effect [Section (2.3)] or with so-called embedded eigenstates, discussed in this section.

The concept of *embedded eigenstates* (EEs), also known as *bound states in the continuum* (BIC) was initially proposed in 1929 by von Neumann and Wigner in quantum mechanics[287]. They constructed a 3D potential extending to infinity and oscillating in a way that the electron wavefunction is predicted to be localized with no radiation within the open system, atom or potential well, inside the continuum of unbounded states. Although in quantum mechanics this concept remains rather a mathematical curiosity[288,289], in recent years, the existence of EEs has been demonstrated in different areas of wave physics, including acoustics[290], hydrodynamics, and photonics[116,291–301] revealing its general nature for wave physics. In a broad sense, EEs are real-valued QNMs of a Hermitian system unbounded from the continuum of radiative modes. In terms of S-matrix poles and zeros, EEs correspond to the presence of a pole at the real frequency axis, where, because of the passivity of the system, it must be canceled out by the respective S-matrix zero. As a result, such a QNM possesses an unboundedly large Q-factor. However, Hermitian systems are a mathematical abstraction, and in reality, such states manifest themselves as quasi-EEs with finite Q-factor and the Fano resonance profile.

Photonics has been demonstrated as a particularly attractive platform for tailoring EEs owing to the existing precise nanofabrication techniques. Photonic EEs have been observed and studied in many open optical systems, including 1D periodic dielectric structures[302–306], 2D photonic crystals[307,308], optical waveguides[295,309–311], quantum-well-based heterostructures[291], and anisotropic birefringent structures[312].

Trivial *symmetry-protected* EEs can be tailored by utilizing symmetries that forbid radiation[116]. However, *non-symmetry-protected* EEs have also been recently predicted and experimentally realized[302,308]. The most crucial feature of EEs is that they behave as discrete states of a closed system within the continuous spectrum of open states and usually manifest themselves as point-like peculiarities in scattering/reflection spectra, see for example Fig. 7(a). This structure proposed in Ref.[119] consists of a dielectric planar resonator sandwiched between two epsilon-near-zero (ENZ) materials with Drude-like dispersion around plasma frequency[313]. If the system is lossless, the scattering spectrum demonstrates EE at the intersection of Fabry-Perot resonance and bulk plasma resonance of ENZ, when the wave impedance for TM waves in ENZ turns to infinity[313,314]. This regime is characterized by divergent grows of Q-factor as one approaches this state by tuning frequency or angle of incidence; see the reflection spectrum in Fig. 7(a). Thus, the EE concept allows tailoring resonances in open systems with *divergently large Q-factor*, corresponding to a scattering pole arbitrarily close to the real frequency axis, namely, with vanishing radiation leakage. If the EE with infinite Q-factor is excited somehow (e.g., with an internal source, or externally by inelastic scattering), this state will live for a very long time, theoretically forever, giving rise to the perfect light-trapping.

> The EE concept allows tailoring resonances in open systems with divergently large Q-factor.

At the same time, the S-matrix analysis shows that tuning the system towards an EE leads to moving of the S-matrix pole and zero closer to each other until they get coalesced at the real frequency axis[96]. We note, however, that the *pole cannot get to the real axis in a passive system* with no gain; otherwise, it would imply lasing. However, there is no limit on how close the pole (and zero) can get to the real axis (and to each other). Hence, in the EE regime, the reflection coefficient or scattering cross-section becomes undefined, revealing the fact that the S-matrix approach becomes inapplicable to the analysis of EEs.

On the contrary, a small detuning from the ideal EE regime establishes an extremely narrow Fano resonance[98]. In other words, open systems supporting ideal EEs provide an ability to localize light without limits and thus enhance light-matter interaction processes. This unique property causes various practical applications of EEs such as lasers[233,235], frequency generation[315], and many others[234].

In reality, no lossless materials and perfect fabrication approaches exist. These imperfections lead to a turning of EE into very narrow Fano resonance (*quasi-EE*) with a high Q-factor [considering the notes regarding the Q-factor definition made in Section (1)]. For example, Fig. 7(b) demonstrates the experimental realization of quasi EE in a photonic crystal made of $Si_3N_4$ with periodically drilled holes[308]. The measurements of the reflection coefficient from this system show a fast narrowing of the resonance around 35° of incidence angle, where the *radiative* Q-factor reaches $10^6$ (demonstrated total Q was limited to $10^4$ due to non-radiative $Q_{nr}$ part caused by material absorption and disordering). Thus, imperfection of the real structure turns the EE into the narrow Fano resonance.

The recent analysis of the asymmetry of a system regarding substrate/superstrate has shown that this asymmetry can also lead to the transformation of EEs into ordinary resonant states[299]. EEs in periodic structures possess a topological nature, which has been revealed recently[316–318]. These topological properties may lead to robustness of the EE against structural impurities, and observation of their nature even in the presence of disorder and imperfections.

It is instructive to consider the Fabry-Pérot mechanism of the EE formation[116]. This mechanism is quite general and considers at least two coupled oscillators with equal eigenfrequencies $\omega_0$ [in contrast to the single mode-based EE shown in Fig. 7(a), where the role of the second oscillator plays the ENZ bulk plasma resonance]. These oscillators are coupled to each other with the coupling strength $\kappa$ (real-valued for near-field coupling), so that the Hamiltonian of this resulting scatterer is $\mathcal{H}_0 = \begin{pmatrix} \omega_1 & \kappa \\ \kappa & \omega_2 \end{pmatrix}$. The essential point here is that the isolated scatterer has no neither dissipative nor radiative losses (Hermitian). When the oscillators are coupled to a waveguide with coupling rate $\gamma$, thereby forming a two-port system [Fig. 7(c), upper panel], the total *Hamiltonian of the scatterer* (two coupled oscillators) becomes

$$\mathcal{H} = \begin{pmatrix} \omega_0 & \kappa \\ \kappa & \omega_0 \end{pmatrix} - i\gamma \begin{pmatrix} 1 & e^{i\psi} \\ e^{i\psi} & 1 \end{pmatrix}, \tag{31}$$

with the eigenstates $\omega_\pm = \omega_0 \pm \kappa - i\gamma(1 \pm e^{i\psi})$. These eigenstates can become real-valued with an appropriate choice of phase retardation function $\psi = kd$, where $d$ is the mutual distance between resonators and $k$ is the wavenumber in the waveguide. Note that the $d = 0$ case when both resonant modes sit at the same point gives rise to the so-called *Friedrich–Wintgen BIC scenario*[234]. In the model Eq. (31), the EEs can be achieved for $e^{i\psi} = \mp 1$ [Fig. 7(c), lower row], which corresponds to the well-known case of Fabry–Pérot EEs scenario. Note, however, that this model does not take into account the dependence of mode coupling on their relative distance, which can become complex in the far-field.

Since no lossless materials exist, it raises a question about loss compensation with material gain. The material loss can be compensated with material gain in the $\mathcal{PT}$-symmetric structures, discussed in the previous section. The case of $\mathcal{PT}$-symmetry EEs has been theoretically discussed in several recent papers by S. Longhi and others[114,115,319] in a system of coupled waveguides and oscillators. We note that in this case of *active* systems, the requirement of EE to be real-valued is not enough because in such systems a QNM can radiate being real-valued, compensating its energy leakage from the gain (laser). Moreover, special attention should be paid to the system *stability analysis*. Namely, no S-matrix poles should be in the upper complex frequency plane, which would make the system unstable (the exponential growth in time).

Different light-matter interaction processes like spontaneous emission, strong coupling, etc. depend not only on Q-factor but also mode volume V. Therefore, despite EEs in infinite structures may provide unbounded Q factors, their large modal volume implies limited interaction enhancements[119]. Thus, the realization of *finite* structures supporting EEs is highly desirable. It can be rigorously shown that for a finite 3D structure to support a truly bounded EE, the structure must be covered by a material with extreme parameters, such as $\varepsilon = \pm\infty$, $\mu = \pm\infty$, $\varepsilon = 0$, or $\mu = 0$ [96,117,234,262]. Finite structures supporting EEs in the form of layered particles with *epsilon-near-zero* (ENZ) shells[96,117,262] and more complicated metamaterial systems[320] have been proposed. However, realistic ENZ phenomena are typically associated with dissipation losses[321], limiting this approach in optics.

Finally, beyond EEs, there are several other nonradiative current configurations in electromagnetically open systems that have been discovered before the advent of photonics[119,322,323]. One such example is an ensemble of radially oriented oscillating dipoles, homogeneously distributed over a sphere, which does not radiate due to symmetry. In the terminology of modern electrodynamics, these states can be referred to as trivial symmetry-protected EEs. We also note that despite in the modern literature the "anapole state" [see Section (5.4)] is often termed a nonradiating state [324], it is rather a *nonscattering state* since it is a solution to Maxwell's equations sustained by an incoming wave [325].

## 5. Scattering by finite objects

### 5.1. Cancellation and enhancement of scattering

The above examples of unusual scattering effects are mostly related to systems whose interaction with external waves can be limited to a few channels - either one or two for a planar system or a waveguide. Particles of finite dimensions represent another important class of scatterers. It is important to understand that, in contrast to planar systems, a simple one- or two-channel treatment of a plane wave scattering by a finite obstacle breaks down, because the scattering of a plane wave by any such system involves *an infinite number of radiation channels*, specifically, the whole basis of spherical harmonics. For this reason, one cannot talk about reflection or transmission in this case. Instead, the scattering cross-section (SCS) becomes a relevant measure of the scattering process. The total scattering cross-section of a bounded object can be expressed as a weighted sum of the scattering contributions of different orthogonal vector spherical harmonics, defined outside the smallest sphere enclosing the three-dimensional scatterer. For a spherically-symmetric object, the SCS can be written as[87]

$$SCS = \frac{\lambda_0^2}{2\pi} \sum_{l=1}^{\infty} (2l+1)\left(|c_l^{TM}|^2 + |c_l^{TE}|^2\right), \tag{32}$$

where $\lambda_0$ is the wavelength in the surrounding medium, and $c_l^{TM}$ and $c_l^{TE}$ are the so-called scattering coefficients or Mie coefficients, which can be calculated from Mie theory in the case of spherically-symmetric scatterers[87]. For a passive structure, $|c_l^{TM,TE}|^2 \leq 1$. These coefficients indicate the contribution of the different scattering channels, and the superscript TM or TE indicates whether the vector spherical harmonic has magnetic or electric field orthogonal to the radial direction.

Not only all channels are involved in the scattering of a plane wave by a finite particle, but a plane wave itself is not an incident waveform in the strict sense. A plane wave can be expanded in *standing* spherical waves – in contrast to the propagating spherical waves in Eq. (5) – according to the well-known partial-wave expansion[65]

$$e^{ikz} = e^{ikr\cos\theta} = \sum_{l=0}^{\infty} (2l+1) i^n j_l(kr) P_l(\cos\theta), \tag{33}$$

where $j_l$ is the spherical Bessel function of order n and $P_l$ is the $l$-th Legendre polynomial [while spherical Hankel functions of the first and second kind, $h_l^{(1,2)}$, represent incoming and outgoing spherical waves, $j_l$ represents a spherical standing wave, and $2j_l(z) = h_l^{(1)}(z) + h_l^{(2)}(z)$]. As a result, the scattering coefficients $c_l^{TM}$ and $c_l^{TE}$ in Eq. (32) are different from the elements of the scattering matrix defined in Eq. (6). Thus, while the scattering matrix elements relate incoming and outgoing spherical waves, the scattering coefficients in Eq. (32) *relate outgoing spherical waves to a standing-wave excitation*. In the relevant case of spherically-symmetric objects (the scattering process preserves the angular momentum $l$ of an incident spherical wave), the scattering coefficients $c_l$ are related to the elements $\lambda_l$ of the diagonal scattering matrix as follows:

$$c_l = (\lambda_l e^{-2ik_0 a} - 1)/2, \tag{34}$$

where $a$ is the outer radius of the spherical scatterer (the coefficients $\lambda_l$ can also be calculated as the reflection coefficients of the $l$-th spherical-transmission-line model of the scattering problem[326]). Thus, while the eigenvalues of the scattering matrix always have unitary amplitude for lossless scatterers at real frequencies (they represent a phase shift, as discussed above), the scattering coefficients may equal zero at real frequencies when an incident propagating plane wave does not produce a scattered spherical wave of a given order $l$.

## 5.2. Invisibility and Cloaking

Invisibility represents a quintessential example of scattering anomaly enabled by engineered metamaterials, and a large body of literature has been produced on this topic. These states correspond to the presence of a scattering zero [should not be confused with S-matrix zero, see Section (1.4)] on the real frequency axis. Consider the general problem of electromagnetic scattering by an arbitrary three-dimensional bounded object (or collection of objects) that is passive. As sketched in Fig. 8(a), the scatterer is illuminated by a monochromatic plane wave propagating in a certain direction. A detector (D) placed downstream from the scatterer in the forward direction will obviously receive less power than in the case of free-space propagation without obstacles, because the incident energy has been *scattered* (i.e., re-radiated in different directions) and/or *absorbed* (i.e., converted into other energy forms, e.g., heat) by the scatterer. This general process of "removing" energy from an incident beam is called *extinction*. Since the energy received by the detector is different in the cases with and without scatterer, the field at the detector location $\mathbf{E}(\mathbf{r}_D)$ must also be different from the incident field $\mathbf{E}_i(\mathbf{r}_D)$, indicating that the forward scattered field $\mathbf{E}_s(\mathbf{r}_D) = \mathbf{E}(\mathbf{r}_D) - \mathbf{E}_i(\mathbf{r}_D)$ is intimately related to the extinction process. Indeed, the relation between the forward scattering and extinction – known as the *optical theorem* – is one of the oldest and most important results of the wave-scattering theory, and it applies to both electromagnetic, acoustic, and matter-wave scattering[87,327]. This theorem states that the extinction cross-section of an object, ECS (in units of area; equal to the total power "extinguished" from the incident wave divided by the incident power density) is proportional to the forward scattering through the relation (written here in scalar form):

$$ECS = \frac{4\pi}{k_0^2} \text{Im}[S(0)], \qquad (35)$$

where $k_0$ is the wavenumber in the background medium, and $S(0)$ is the forward scattering amplitude, which is related to the scattered electric field as $E_s = \frac{e^{ik_0 r}}{r} S$ [16] if the observation point (the detector) is sufficiently far from the scatterer ($k_0 r \gg 1$). To make an object truly invisible, it is, therefore, necessary to completely cancel its total extinction cross-section, and therefore the forward scattering, such that a detector in the forward direction would receive the same power as in the case with no scatterer. Perfect invisibility is therefore clearly impossible in the case of an absorbing object, as part of the incident energy is unavoidably lost, hence producing a shadow that reveals the presence of the object. In the lossless scenario, the extinction cross-section is equal to the scattering cross-section, SCS, and ideal invisibility is, in principle, possible.

As noted above, some of the scattering coefficients of a finite object may equal zero at real frequencies. According to Eq. (34), this occurs when the scattering process introduces a phase shift $s_l = e^{2ik_0 a}$ (precisely equal to the phase delay acquired by a spherical wave in a spherical region of empty space of radius $a$), whereas maximum *resonant* scattering occurs when the phase shift equals $e^{i(2k_0 a + \pi)}$. As the frequency is varied, the scattering coefficients of any object exhibit an alternation of zeros and maxima. As an illustrative example, Fig. 8(b) shows the magnitude of the first TM scattering coefficient $|c_1^{TM}|$ which corresponds to electric dipolar radiation[16], for a dielectric sphere with radius $a$ and permittivity $\varepsilon = 5$, illuminated by a propagating plane wave. It is therefore clear that the zeros of the scattering coefficients do not correspond to perfect coherent absorption (the scatterer is lossless) but to zero energy scattered into a given channel (spherical harmonic), under plane-wave excitation. The first naturally-occurring zero is sometimes interpreted in terms of a so-called *anapolar* distribution of polarization current with zero dipolar scatterings (see, e.g.,[328]), which is the subject of the next subsection.

The role of cloaking devices is to introduce additional scattering-coefficient zeros at the desired frequencies. More in general, an invisibility cloak aims at reducing the total scattering cross-section Eq. (32) by minimizing all the non-negligible scattering coefficients in the desired frequency window, such that the incident field is restored, $\mathbf{E} = \mathbf{E}_i$, at any point around the cloaked object, and especially in the forward direction. Different techniques exist nowadays to achieve invisibility, including plasmonic and mantle cloaking[329,330], transformation-optics cloaking[331,332], transmission-line cloaking[333], among many other techniques, and we refer an interested reader to topical reviews (e.g., [91]) for detailed information on this broad area of research. Here, instead, we focus on a specific case to give the reader some physical insight on how scattering-cancellation cloaking works, and on the general limits of passive cloaking.

Consider the dielectric sphere in Fig. 8(b), surrounded by a concentric layer made of a different non-magnetic material with permittivity $\varepsilon_c$ and outer radius $a_c$. If the incident wavelength is much longer than the scatterer size, the scattering process is typically dominated by the response of an electric dipole induced on the object by the incident field[87]. In this so-called *quasi-static* regime, the Mie coefficients can be greatly simplified using a small-argument approximation, and the conditions to achieve invisibility assume a simple algebraic form. Notably, the condition to suppress the dominant scattering coefficient, $c_1^{TM} = 0$, becomes[329]

$$\left(\frac{a}{a_c}\right)^3 = \frac{(\varepsilon_c - \varepsilon_0)(2\varepsilon_c + \varepsilon)}{(\varepsilon_c - \varepsilon)(2\varepsilon_c + \varepsilon_0)}. \tag{36}$$

This condition can be satisfied only if the right-hand-side is positive and smaller than one, which implies that for a dielectric sphere with $\varepsilon > \varepsilon_0$, the cloaking shell must have $\varepsilon_c < \varepsilon_0$. A shell of this type can be realized with plasma-like materials, such as noble metals at optical frequencies[87] or wire-based metamaterials at microwave frequencies[334,335], hence the name "plasmonic cloaking"[329]. Interestingly, because of the quasi-static nature of this analysis, the incident-wave frequency does not explicitly appear in Eq. (36); however, a material with $\varepsilon < \varepsilon_0$ must always have frequency-dependent permittivity due to passivity and causality considerations[336]. As a result, no matter the frequency-dispersion model of the considered material (e.g., Drude or Lorentz model[336]), condition Eq. (36) can be exactly satisfied only at a single frequency (or a discrete set of frequencies), producing a minimum in the scattering spectrum, as shown in the example in Fig. 8(c). Around this minimum, the scattering cross-section remains moderately low over a narrow frequency bandwidth, but the cloaking performance rapidly deteriorates away from the central frequency. In fact, the scattering cross-section of the cloaked object may actually become larger than that of the bare object at nearby frequencies, as seen in Fig. 8(c). As recently investigated in different works[337–340], the problem of narrow bandwidth represents a general issue for passive invisibility devices, independent of the employed cloaking technique. Passivity and causality determine fundamental limits on the achievable level of *continuous* scattering reduction over a given bandwidth (perfect invisibility can be achieved only at individual frequencies[341]). Fig. 8(e) shows an illustrative example of physical bound on passive cloaking for a spherical object of a given radius, and we refer the reader to Ref.[337] for further details.

In addition to these "local" bounds on the performance of cloaking devices in a given frequency window, passivity and causality also determine "global" bounds on the response of invisibility cloaks over extensive frequency ranges, as discussed in Ref.[342]. These global cloaking bounds are based on an interesting sum rule derived by E. M. Purcell in 1969 [343], which allows relating the total ECS (or the SCS in the lossless case) of a given object, integrated over the entire electromagnetic spectrum, to the *static* properties of that object represented by its static polarizability. For a nonmagnetic lossless object with static polarizability tensor $\alpha_{e,s}$, this sum rule reads [342]

$$\int_0^\infty SCS\, d\lambda = \pi^2 (\hat{p}_e^* \cdot \alpha_{e,s} \cdot \hat{p}_e), \tag{37}$$

where $\lambda$ and $\hat{p}_e$ indicate the wavelength and the polarization unit vector of the incident wave. For example, for a dielectric nanosphere having static permittivity $\varepsilon_s = 15\varepsilon_0$ and radius a=100 nm, the right-hand-side of Eq. (37) is equal to $\pi^2 \alpha_{e,s} = 4\pi^3 a^3 (\varepsilon_s - \varepsilon_0)/(\varepsilon_s + 2\varepsilon_0) \approx 0.01$ $\mu$m$^3$. Interestingly, it was demonstrated in Ref.[342] that, if we surround the considered object with an arbitrary passive shell – no matter its complexity – this number is always bound to increase. This finding implies that, while an invisibility cloak can be designed to efficiently suppress the scattering cross-section of an object in a given frequency window, the scattered energy will necessarily increase in other regions of the electromagnetic spectrum. In other words, a *passive* *cloaked object always scatters more*, not less, than the original object, when illuminated by an electromagnetic signal of sufficiently wide bandwidth.

> A passive cloaked object always scatters more, not less, than the original object, when illuminated by an electromagnetic signal of sufficiently wide bandwidth.

Both global and local bounds on cloaking can be relaxed if the background medium surrounding the object to be concealed is modified, for example, if the cloaked object is embedded in a high-index dielectric[344] or in a diffusive light-scattering medium (e.g., fog)[345]. In addition, relaxed versions of invisibility may allow broader operational bandwidth, especially if only a discrete set of incident/detection angles are of interest, and/or if one aims at restoring only the amplitude (not the phase) of the incident field around the cloaked object (see, e.g., [346–349]).

Finally, new opportunities are offered by active scattering systems, which may overcome the conditions imposed by passivity that underpin most of the limits discussed above. For example, it may be possible to broaden the bandwidth of cloaking devices by including active elements in their design[350]; however, the *stability* of these active scattering systems should be carefully assessed, to avoid self-oscillations and amplification effects that would make the cloaked object significantly more visible. Furthermore, active systems with suitable combinations of gain and loss (for example $\mathcal{PT}$-symmetric scatterers) may be designed to achieve perfect invisibility while they absorb a portion of the impinging energy. This is possible thanks to the fact that, in active systems, the optical theorem Eq. (35) can be satisfied with zero forward scattering *and* finite absorption provided that the effect of gain exactly compensates losses, hence suppressing the ECS. This approach has been used to realize invisible cloaked sensors for acoustic waves[351] and unidirectional electromagnetic cloaks for large objects[258], as shown in Fig. 8(f). In addition to the stability issues described above, an important challenge for $\mathcal{PT}$-symmetric invisible scatterers is that a balanced gain-loss distribution is typically achieved in a specific direction, whereas either gain or loss dominates in other directions. To achieve invisibility for incident waves impinging from multiple directions, it may be necessary to make the active scattering system *spatially-dispersive*, namely, angle-dependent, as recently proposed in the context of $\mathcal{PT}$-symmetric metasurfaces for focusing and imaging[352].

In general, new approaches based on active, nonlinear, nonreciprocal, nonlocal systems are expected to open new exciting directions in the science of invisibility and break the limitations of conventional stealth technology.

## 5.3. Superscattering

As shown in Section (1), QNMs couple to different channels not equivalently, i.e., with different coupling strengths. For example, if a scatterer QNM can be excited by monochromatic excitation through only one channel, it will scatter to the same channel only owing to the reciprocity principle.

In general, the choice of a particular set of channels is dictated by the symmetry. For example, for spherical or cylindrical objects the basis of spherical and cylindrical waves are naturally dictated channels. In this case of highly symmetric structures, the scattering matrix $\hat{S}$ is reduced to its diagonal form $\hat{S}_D = \text{diag}(d_1, d_2, \cdots)$, where $d_l$ is the scattering amplitude to the same channel. This means that excitation of a spherical (cylindrical) scatterer by a spherical (cylindrical) partial monochromatic wave will cause scattering to the same channel. It is worth mentioning that any transient signal has a finite spectrum and hence the partial wave can excite other modes that hit this spectrum.

If the system is passive, the energy conservation law requires the reflection coefficient to the same channel to be $|S_l|^2 \leq 1$. This energy conservation law also imposes the limit of scattered energy to *one channel* when the structure is excited through several channels. For example, under a plane wave excitation, the maximum scattering cross-section of light at the resonance of 3D scatterer with the total angular momentum $l$ is found to be $(2l+1)\lambda^2/2\pi$, which gives for a dipole resonance ($l=0$) the value of $3\lambda^2/2\pi$ [353]. Fundamentally, this limit stems from the energy conservation law and the fact that a plane wave excitation is equivalent to incoming energy in all possible channels. This restriction also stems from Eq. (32) for the case $|c_l^{TM,TE}|^2 = 1$.

However, one can get around this limit by employing multiple QNMs with different total angular momenta and ensuring that all these QNMs have the same real frequency (the imaginary frequency may be different) and hence can interfere constructively. Indeed, a plane wave excitation because of its multi-channel nature, allows different modes to be excited, and the scattered energy thus can go into more than one channel, overcoming the one channel limit. This effect is called as *superscattering* regime[354] and has been proposed for scatterers of various geometries, including cylinders[354–357], spheres[358–361], double-slit structures[362], nanodisks and has been demonstrated experimentally[363].

We note that the superscattering effect is opposite to the coupled resonator *induced transparency*[364–366] when scattering from different QNMs destructively interfere in the far-field and hence reduce overall scattering rather than increase it[367–370]. It has been also demonstrated that the same structure could be designed to have both regimes in the same frequency range[361,362] such that it can leap from induced transparency to superscattering and back through structural parameters variation. We also note that the superscattering effect is also related to the effect of *superdirectivity* in antenna theory[176,371–374], where the excitation of several different multipoles in subwavelength antennas at the same frequency leads to highly directive emission.

## 5.4. Anapole

The last type of unusual scattering phenomena that we consider here is the so-called *anapole state*. In a broad sense, the anapole is a spatial distribution of oscillating currents that do not scatter an electromagnetic field[328,375–382]. Although, the *anapole may be seen as a particular case of* *cloaking* realized when the scatterer has a zero of scattering amplitude (not an S-matrix) on the real axis, in the modern literature the two phenomena are commonly distinguished, since they pursue different goals, namely to make an object invisible (cloaking) and to enhance the internal fields (anapole)[378].

In the context of classical electrodynamics, an anapole resonant state corresponds to a specific type of charge-current distribution that does not scatter. However, this fact does not mean that anapoles do not interact with external fields. In this context, we note that in the literature, the anapole is sometimes referred to as a (nonradiative) mode. However, it is important to stress that the *anapole is not a natural QNM* of an open cavity constituting the scattering object, or in other words, *cannot support self-sustained oscillations*. It is, instead, a distribution of fields (or polarization currents) that can be excited by a suitable impinging field distribution. It just happens that this combination of polarization currents does not sustain a scattered wave, like other nonscattering field distributions extensively studied in the context of invisible or "cloaked" bodies, see Section (5.2). The non-scattering nature of the anapole stems from the destructive interference of waves emitted by several constituent QNMs. In the result of such interference, one of the zeros of scattering amplitude can get to the real frequency axis even in the case of a lossless system[86].

The drastic difference between anapole and truly nonradiative QNM (embedded eigenstates in the continuum, EE) appears in the transient regime after turning off the external excitation. In this case, the anapole state will decay, whereas the EE won't radiate, being a self-sustained eigenmode of the structure. This behavior is related to the implications of Lorentz reciprocity for nonradiating current distributions[325].

In general, *any* spatial charge-current distribution regardless of the size and shape of an object can be expanded into a series of *multipole moments*[16]. Unlike QNMs, multipole moments depend on the choice of orthogonal functions (e.g., spherical harmonics, plane waves), used as a basis. On the other hand, w*hen the scatterer is small enough compared to the wavelength*, a Taylor expansion in Cartesian coordinates can be applied for scattering analysis. In addition to electric and magnetic dipoles, quadrupoles, etc., the expansion in Cartesian coordinates allows calculating the toroidal moments[383,384], which are higher-order Taylor expansion contributing to lower-order spherical harmonics. In this sense, an anapole source is a superposition of two elementary multipoles: Cartesian electric dipole (ED) and Cartesian toroidal dipole (TD) of a certain amplitude oscillating in antiphase (to fulfill the destructive interference requirement)[385–387], Fig. 9(a). An ED source is merely a linear oscillating current, while a spatial structure of a TD source is a bit more sophisticated. The TD is formed by circulating electrical currents flowing on a surface of an infinitesimal torus. The resulting current distribution is associated with a TD moment pointing outward along the torus symmetry axis. The TD of a given current distribution can be calculated according to (in SI units)[384,388]:

$$\mathbf{T} = \frac{1}{10c}\int [(\mathbf{r}\cdot\mathbf{J})\mathbf{r} - 2r^2\mathbf{J}]d^3\mathbf{r}, \qquad (36)$$

where **J** is the induced electric current density related to the local electric field *via* $\mathbf{J} = -i\omega\varepsilon_0(\varepsilon-1)\mathbf{E}$ with $\varepsilon$ being local permittivity. The corresponding electric field in the far-field of the TD is[323,375,388]

$$\mathbf{E}_{\mathrm{TD}} = ik\frac{k^2}{4\pi\varepsilon_0}\mathbf{n}\times\mathbf{T}\times\mathbf{n}, \tag{37}$$

where **n** is the unit vector in the observation direction. A special feature of the TD is that its far-field pattern is identical to that of the ED, which has the form $\mathbf{E}_{\mathrm{ED}} = \frac{k^2}{4\pi\varepsilon_0}\mathbf{n}\times\mathbf{P}\times\mathbf{n}$. We immediately recognize that when the ED and TD satisfy the relation $\mathbf{P} = -ik\mathbf{T}$, which defines the anapole condition, their far-field radiation vanishes, making such a source nonscattering. As a result, the anapole manifests itself as the emergence of a scattering zero in the electric dipole spherical harmonic channel (Fig. 9b; see also Fig. 8b).

Nevertheless, the *ED and TD are distinguishable within the scatterer*, where one may perform a Cartesian multipole expansion of the induced currents and calculate the induced ED and TD[389]. For a certain geometry of the scatterer and incident wavelength, the incident wave may excite ED and TD satisfying $\mathbf{P} = -ik\mathbf{T}$ in the particle forming the anapole configuration. Fig. 9(b) shows the absolute value of the $c_1^{\mathrm{TM}}$ Mie coefficient (often denoted as $a_1$) for a spherical particle as a function of wavelength. At a specific wavelength $c_1^{\mathrm{TM}}$ turns to zero, indicating ideal destructive interference between the induced ED and TD in the far-field[328]. The same situation may take place for a cylinder illuminated by a plane wave[386]. Notably, the $d_1^{\mathrm{TM}}$ coefficient, which reflects the amplitude of the $\mathrm{TM}_1$ spherical harmonic *inside* the particle, does not vanish at this point – the $\mathrm{TM}_1$ harmonic inside the particle representing the sum of ED and TD is not zero, but it is their combination that does not scatter.

As one can see in the anapole currents in a spherical particle are excited away from the resonance, Fig. 9(b). This situation corresponds to the formation of a particular current configuration in the particle that does not scatter in the $\mathrm{TM}_1$ (ED) channel. However, apart from the anapole, the incident wave will excite all other modes of the particle that will scatter light in other channels. A nanoparticle can be made to scatter less by employing a special incident beam[377], which excites mostly the anapole, but such a scattering suppression still won't be exact. Nevertheless, it is possible to construct such a potential (geometry and permittivity of a nanoparticle), in which a certain eigenmode will be dominantly characterized by TD and ED[390], which cancel each other in the far-field. An example is a dielectric nanodisk with a large diameter/height ratio. The electric field distribution inside the nanodisk at the resonance wavelength, Fig. 9(c), clearly indicates the formation of an anapole-like field distribution. At higher frequencies, higher-order anapole states can be found with more complicated field distribution[391]. Surprisingly, such an anapole-like state manifests itself as an anti-resonance in the scattering spectrum, opposed to the expected peak, Fig. 9(c). The internal field, on the other hand, is enhanced at the anapole wavelength, together with the prominent enhancement of the Purcell factor, Fig. 9(c).

Resonant features of large aspect ratio dielectric nanodisks associated with anapole-like states may be utilized to enhance light-matter interaction. In particular, they have been employed for the enhancement of nonlinear effects, including third-harmonic generation and four-wave mixing, Fig. 9(d)[92–95].

The last question we touch here is the recently suggested concept of the so-called "*anapole nanolaser*"[392]. It is known, that the rigorous laser theory requires a system to have a well-defined mode with a pole (of scattering or S-matrix eigenvalue) in the complex plane and exact total loss compensation when this pole reaches the real frequency axis leading to the self-sustaining lasing regime[33,140]. Hoverer, since an anapole is a zero, not a pole, of the scattering coefficient, an anapole state *cannot be used directly to achieve lasing*.

## Conclusions

In this paper, we have reviewed and put in a unified context a variety of anomalous scattering effects in nanophotonics, showing the intricate relation between them. We established a general framework, based on the analytical properties of the scattering matrix, to look at exotic and often counter-intuitive scattering phenomena from the same ground and develop a unified picture for these nanophotonic phenomena. We have shown that perfect absorption, as well as coherent and virtual absorption, the interplay of gain and loss, PT-symmetry breaking, exceptional points, bound states in the continuum, cloaking, and anapoles can be all interpreted in terms of the evolution of the poles and zeros of the scattering matrix. We summarize this in Table (1). Our presentation shows that manipulation of these relevant points in the scattering of a system opens disruptive opportunities for a variety of applications, from integrated photonic systems to sensing, computing, and analog signal processing. Arrays of these exotic scatterers also open relevant opportunities in the area of metamaterials and of artificial materials with unusual optical properties.

| **Phenomenon** | **Pole/Zero description** | **Section** |
|---|---|---|
| Laser | Pole of S-matrix *or* scattering coefficient at the real frequency axis (classic definition) | 1.6, 2.4 |
| Perfect absorption (PA) | Zeros of reflection coefficient *and* transmission coefficient at the real frequency axis | 2.1 |
| Coherent perfect absorption (CPA) | Zero of S-matrix at the real frequency axis | 2.2 |
| Virtual perfect absorption (VPA) | Zero of S-matrix in the upper plane; exponentially growing excitation | 2.3 |
| Virtual gain | Passive system; Excitation attenuates faster than the cavity decay rate | 2.3 |
| CPA-laser | Zero and pole of S-matrix coincide at the real frequency axis | 3.2 |
| Zero scattering (Anapole, cloaking, tunneling) | Zero of scattering coefficient at the real frequency axis | 1.4, 5.2, 5.4 |
| Unidirectional invisibility | PT-symmetric structures; Zero of one or more scattering coefficients *and* full transmission at the real frequency axis | 3.3 |
| Exceptional point (EP) | Coalescence of two or more S-matrix poles at the real frequency axis (or in the lower plane) | 3.1 |
| Bound state in the continuum (BIC) | Coalescence of zero and pole of S-matrix at the real frequency axis | 4 |
| BIC-like scattering | Active systems; Coalescence of *scattering zero* and S-matrix pole at the real frequency axis | 1.4 |
| Superscattering | Interferon of radiation from two QNMs; Two poles in the lower plane with the same real frequency | 5.3 |

**Table 1**. Poles and zeros description of different scattering effects discussed in this work.

## Acknowledgments

This work was supported by the Air Force Office of Scientific Research, the Simons Foundation and the National Science Foundation. Authors acknowledge fruitful discussion with Andrey Bogdanov (ITMO University), Andrey Miroshnichenko (University of New South Wales Canberra), and Philippe Lalanne (CNRS-Institut d'Optique Graduate School-Univ. Bordeaux).

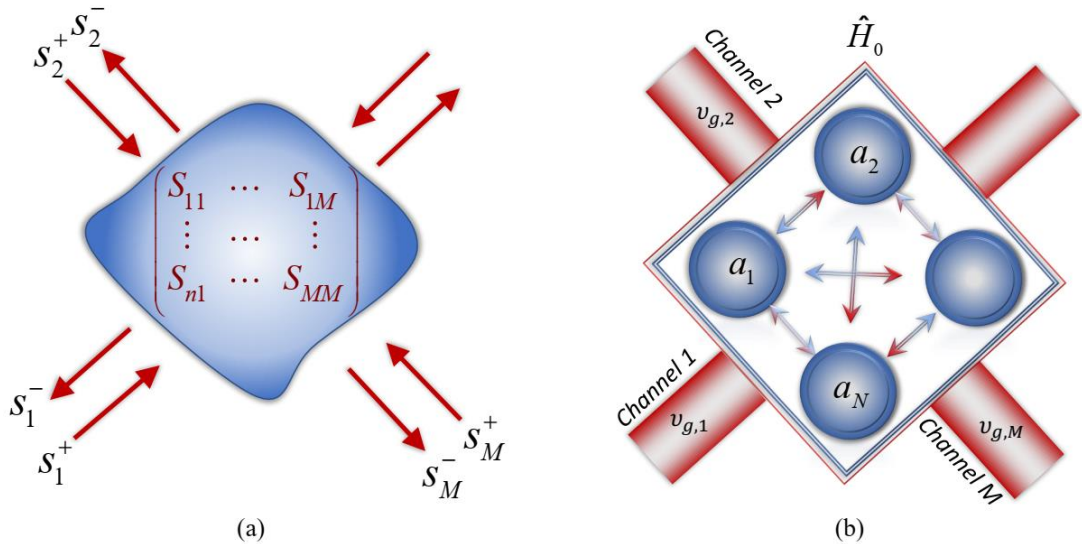

**Figure 1**. (a) Description of a linear electromagnetic structure by its scattering ($\hat{S}$) matrix, which links the amplitudes of ingoing ($s_n^+$) and outgoing ($s_n^-$) waves; $M$ stands for the number of channels. (b) Effective Hamiltonian approach to scattering description. Here $a_n$ stands for the amplitude of the *n*-th mode of the idealized closed system. Due to imperfections like defects or deformations, the modes interact with each other (shown by arrows). Together, they constitute the closed-system Hamiltonian $\hat{H}_0$ of the scatterer. The system becomes open after coupling it to channels.

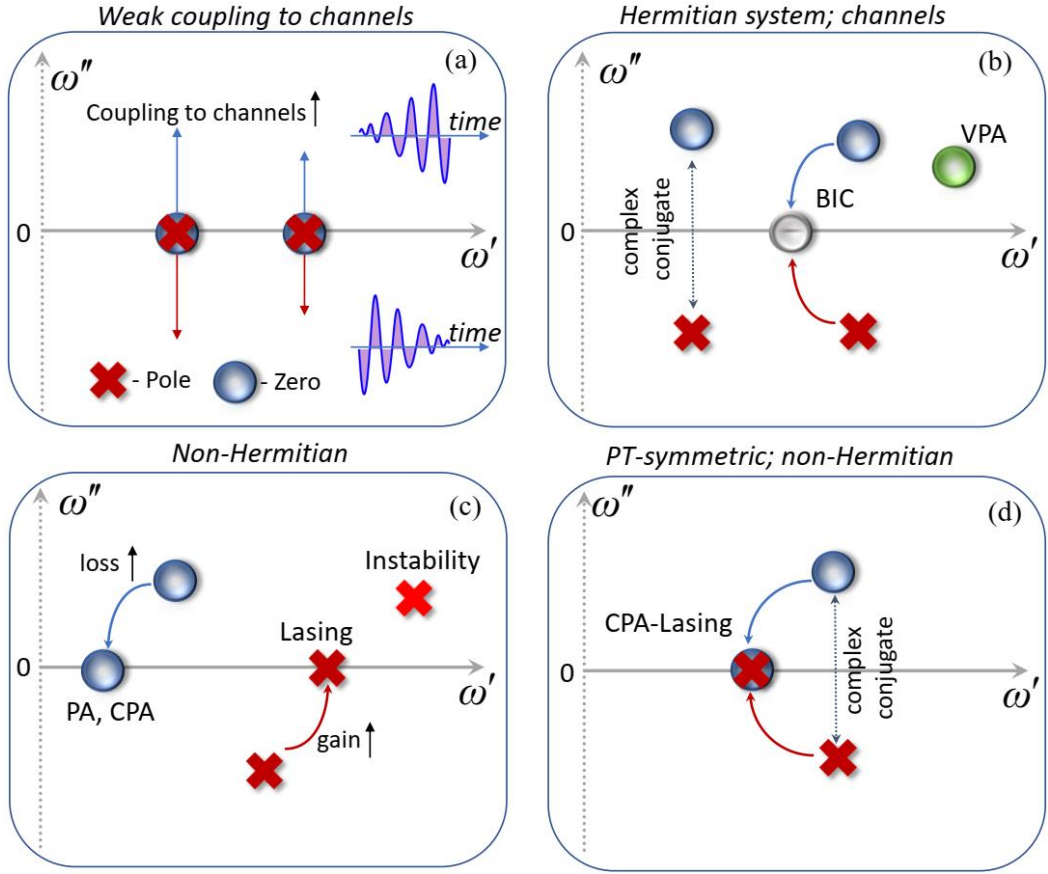

**Figure 2**. **Various scenarios of light scattering and their relation to poles and zeros location.** (a) Hermitian scatterer with weak coupling to channels. In the limit of almost no coupling to channels, poles and zeros lie very close to the real axis. With increased coupling to channels (radiative losses), the pole (zero) goes to the upper (lower) hemispace. (b) General Hermitian scatterer. The zeros and poles are related to each other through complex conjugation. Bounds state in the continuum (BIC) are the limiting case when the zero and pole of a Hermitian system coalesce on the real axis with coherent cancelation (destruction) of each other. VPA stands for Virtual Perfect Absorption. (c) Different scattering scenarios for non-Hermitian scatterer, including lasing, perfect absorption (PA), coherent perfect absorption (CPA). The presence of a pole in the upper plane turns the system into an unstable regime, with eigenmodes that grow to infinity as time evolves. (d) A particular case of non-Hermicity, $\mathcal{PT}$-symmetric scatterers, when material loss and gain are balanced. In this scenario, the CPA-laser regime is possible when the pole and zero coalesce in the same point at real axes in the so-called $\mathcal{PT}$-symmetric phase.

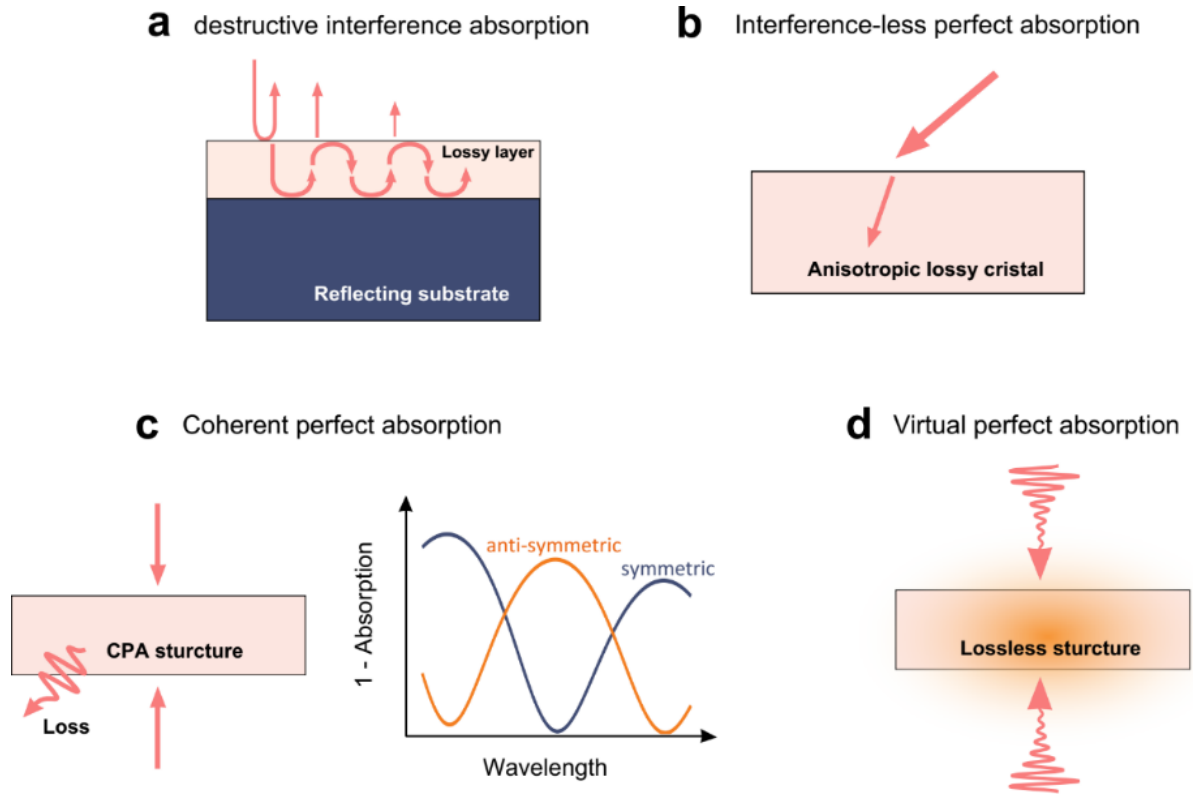

**Figure 3**. **Various scenarios of perfect absorption (PA).** (a) PA in a one-port system enabled via destructive interference. (b) PA in a one-port system enabled via lossy Brewster angle. (c) Left: coherent PA in a two-port system. Right: a typical spectrum of absorptivity for two eigenmodes (even and odd) of a two-port structure. (d) Virtual PA: illumination of a lossless (Hermitian) system with exponentially diverging fields causes the reflection to vanish.

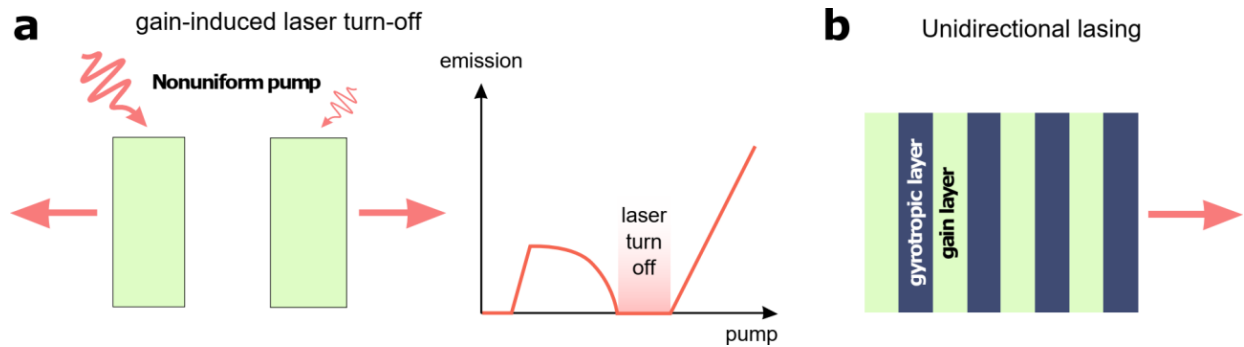

**Figure 4. Unconventional lasing scenarios.** (a) Gain-induced laser turn-off induced by non-uniform pumping of the system. Left: schematics of the structure. Left: typical dependence of the lasing intensity on the pump parameter exhibiting laser turn-off with increasing pump. (b)

Unidirectional lasing from a periodic structure enabled by the magneto-optical response of the constituting layers.

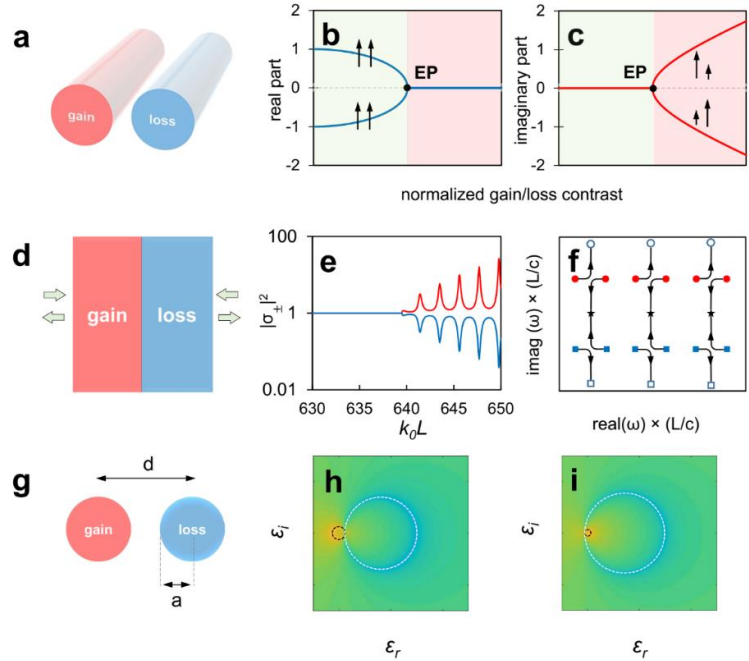

**Figure 5. $\mathcal{PT}$-symmetric structures.** (a) $\mathcal{PT}$-symmetric arrangement of coupled optical waveguides with gain and loss. (b,c) The real and imaginary parts of the normalized propagation constants of the coupled $\mathcal{PT}$ waveguides. $\mathcal{PT}$-symmetric 1D scattering setting composed of two dielectric slabs with equal gain and loss, (e) the absolute value of the scattering matrix eigenvalues $|\sigma_{1,2}|$ versus the electric length $k_0 L$ of the Fabry-Perot cavity. (f) Evolution of the zeros (circles) and poles (squares) of the scattering matrix eigenvalues of the $\mathcal{PT}$ Fabry-Perot in the complex frequency plane for an increasing gain-loss parameter g. (g) $\mathcal{PT}$-symmetric arrangement of two dielectric particles with complex conjugate permittivities. (h,i) Poles and zeros of scattering in the complex permittivity plane for longitudinal and transverse polarizations, respectively. Images reprinted with permissions from: (e, f) – Ref.[253], (h, i) - Ref.[261].

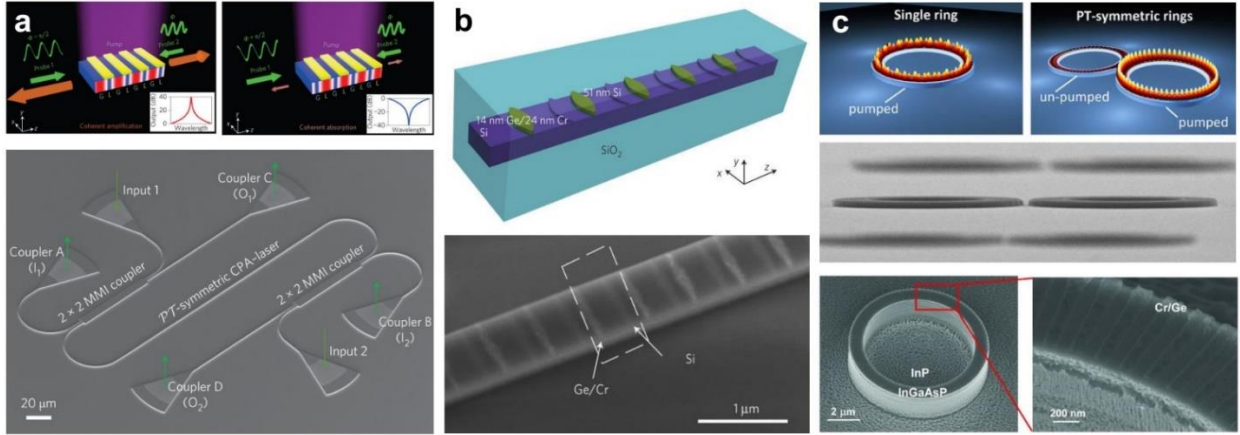

**Figure 6. Applications of $\mathcal{PT}$-symmetric structures.** (a) Experimental demonstration of simultaneous lasing and anti-lasing. Lasing (left) and absorber (right) behavior. The panel below demonstrates the SEM image of the fabricated structure. (b) Unidirectional invisibility. Upper panel: schematic picture, lower panel: the SEM image. (c) Single-mode lasing. Images reprinted with permissions from: (a) – Ref.[163], (b) – Ref.[267], (c) – Ref.[241,242].

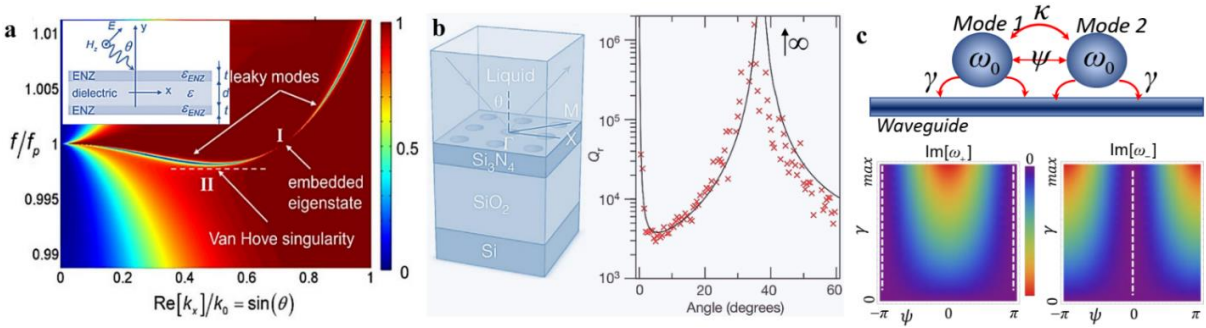

**Figure 7. Bound states in the continuum.** (a) Leaky mode bounded in the continuum. The structure is a waveguide composed of a dielectric slab ($\varepsilon_d = 5$) sandwiched between two ENZ layers with Drude-like dispersion, as shown by the inset. The reflection spectrum demonstrates EE at an intersection of Fabry-Perot resonance and the bulk plasmon resonance ($f_p$) of ENZ slabs. (b) Experimental realization of EE with a photonic crystal $Si_3N_4$ slab with periodically drilled holes. (c) Upper panel: Fabry-Perot model of BICs. Lower row: Imaginary value of both eigenstates as a function on radiative decay rate ($\gamma$) and phase retardation ($\psi = kd$). Vertical dashed lines denote EE positions. Images reprinted with permissions from: (a) – Ref.[119], (b) – Ref.[308].

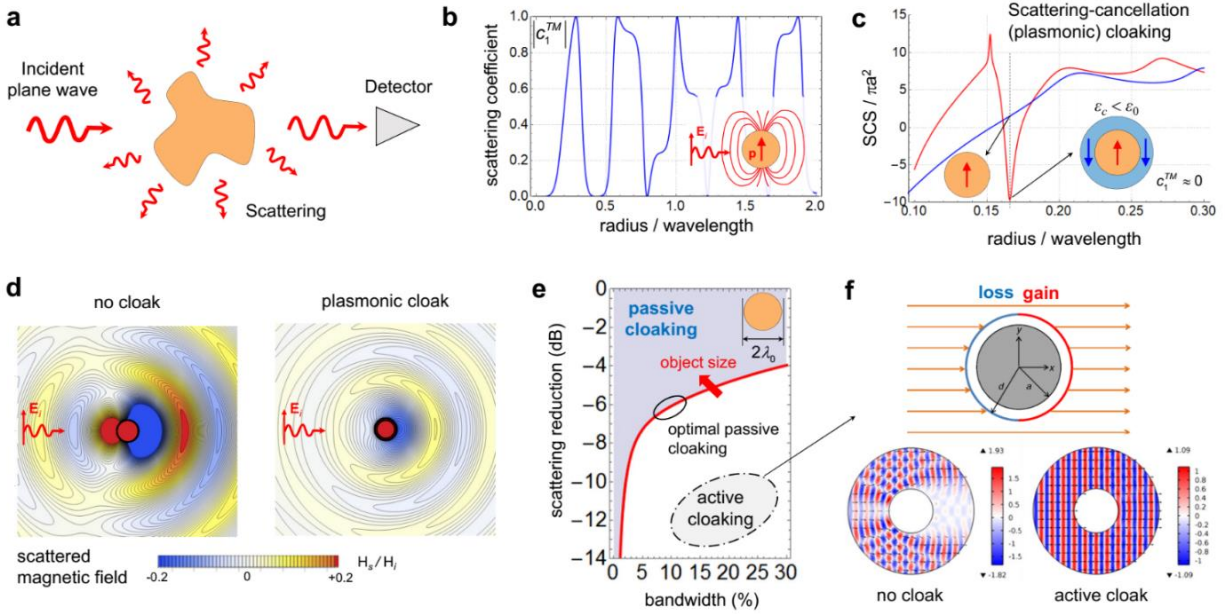

**Figure 8. Invisibility and Cloaking.** (a) General problem of wave scattering by a material body illuminated by a plane wave. A detector in the forward direction receives less energy than in the case without passive obstacles. (b) First transverse-magnetic (TM) scattering coefficient for a homogeneous sphere with permittivity $\varepsilon = 5$, as a function of its radius normalized by the wavelength of the incident wave. This quantity indicates the scattering contribution due to an induced electric dipole, as sketched in the inset. (c) Normalized scattering cross-section of the scatterer in (b) without (blue curve) and with (red) a plasmonic cloak designed with Eq. (29), assuming a silver-like Drude dispersion for the plasmonic material. The effect of the cloak is to minimize the electric dipolar scattering by inducing counter-oscillating dipole moments in the core and shell. (d) Time-snapshot of the out-of-plane scattered magnetic field for the uncloaked (left) and cloaked (right) sphere considered in (c). The plasmonic cloak minimizes the electric dipolar scattering; the residual scattering is due to higher-order scattering contributions (in this case, an out-of-plane magnetic dipole). (e) Physical bound on passive cloaking, indicating the optimal trade-off between scattering reduction and bandwidth for any passive cloak applied to a spherical object of radius $a = \lambda_0$. The bound becomes more stringent as the object size increases. Adapted with permission from[337]. (f) Active $\mathcal{PT}$-symmetric cloak for unidirectional invisibility of large objects. Images reprinted with permissions from: (f) – Ref [258].

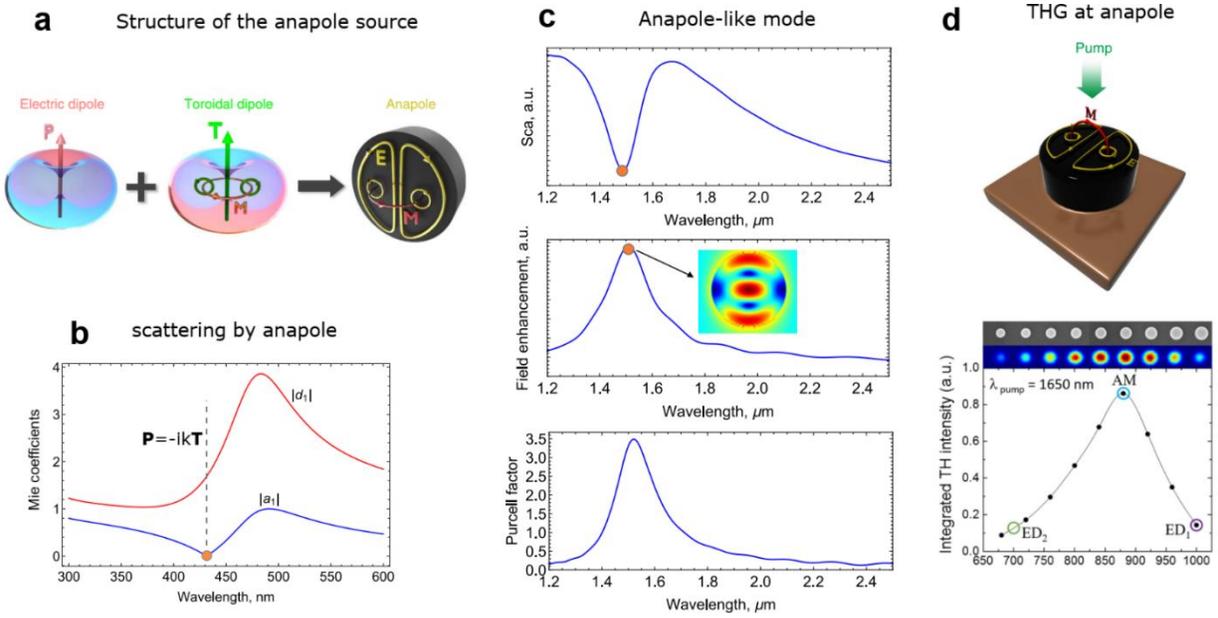

**Figure 9. Anapole.** (a) Structure of an anapole source: a superposition of collinear Cartesian electric dipole and a Cartesian toroidal distribution of electric currents. (b) Manifestation of anapole in the scattering of a plane wave by a dielectric spherical particle of 80 nm radius with $\varepsilon = 16$. The $c_1^{TM}$ Mie coefficient vanishes at a particular frequency, indicating destructive interference of ED and TD in the far-field. The $d_1^{TM}$ coefficient at the same time has a non-zero value responsible for the ED and TD currents inside the particle. (c) Electromagnetic response of a nanodisk supporting anapole-like field distributions. Scattering cross-section (top), averaged electric field enhancement (middle), and the Purcell factor of an electric dipole placed at the center of a disk with refractive index n=4.2 with 100 nm height and 350 nm radius. (d) Bottom: THG enhancement by a nanodisk with the anapole-like mode. The nanodisk supporting anapole-like mode at the incident wavelength exhibits maximal THG signal. Top: Schematic of a nanodisk laser operating at the anapole-like mode. Images reprinted with permissions from: (a) – Ref.[328], (d) – Refs.[92,392].

## About the Authors

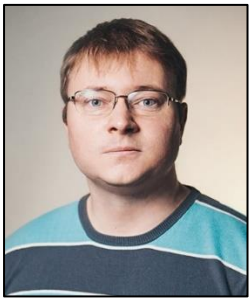

**Alex Krasnok** earned his Ph.D. with distinction from ITMO University (St Petersburg, Russia) in 2013. After spending two years (2016-2018) as a research scientist at The University of Texas at Austin (Austin, USA), he joined the Advanced Science Research Center CUNY (New York, USA) as a Research Assistant Professor and Core Facility Director. His current research interests are in the areas of applied electromagnetics, nanophotonics, metamaterials, plasmonics, and nanotechnology, with particular emphasis on cross-disciplinary research. He has made significant contributions in the areas of extreme scattering engineering, nanoantennas, metasurfaces, optics of 2D transition-metal dichalcogenides, and low-loss dielectric nanostructures. He has earned several research awards, including the gold medal of Nobel Laureate Zhores Alferov's Foundation (2016).

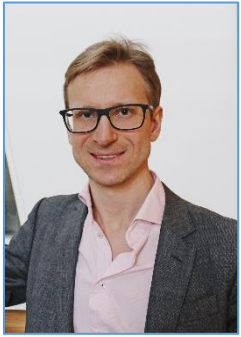

**Denis Baranov** received his Ph.D. degree in physics from Moscow Institute of Physics and Technology in 2016, and now he is a postdoctoral researcher at Chalmers University of Technology. His interests include perfect electromagnetic absorption and lasing, ultrafast and nonlinear interaction of light with resonant nanophotonic structures and extraordinary regimes of strong light-matter interaction.

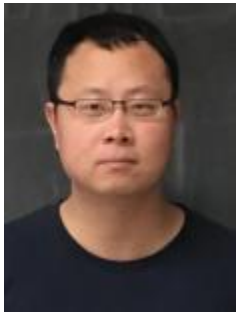

**Huanan Li** is currently a postdoctoral research fellow in the photonics initiative of CUNY Advanced Science Research Center, New York, NY, USA. He received his B. Sc. Degree in physics in 2009 from Sichuan University in China, and then the Ph. D. degree in 2013 from physics department of National University of Singapore, Singapore. From 2014 to 2018, he is working as a post-doctoral research associate in the Wave Transport in Complex Systems Lab of the department of physics in Wesleyan University, Middletown, USA. His works centered between mathematical physics and applied physics. His research interests focus on wave physics and its methodology in non-Hermitian and time-dependent systems. He has published around 30 papers in peer-reviewed scientific journals and one book chapter.

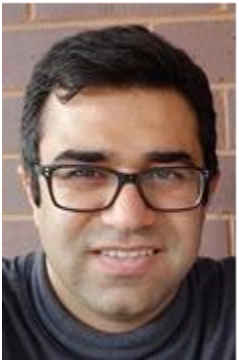

**Mohammad-Ali Miri** is an Assistant Professor of Physics at Queens College and the Graduate Center of the City University of New York. He earned his Ph.D. in Optics from CREOL, the College of Optics and Photonics, at the University of Central Florida in 2014. His research interests are in the broad areas of optics and photonics, light-matter interaction, and nonlinear wave dynamics with a particular emphasis on the physics of complex nonlinear and non-Hermitian systems and their applications in photonics.

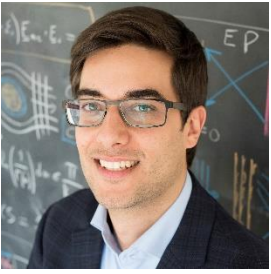

**Francesco Monticone** is an Assistant Professor of Electrical and Computer Engineering at Cornell University. He received the B.Sc. and M.Sc. (summa cum laude) degrees from Politecnico di Torino, Italy, in 2009 and 2011, respectively, and the Ph.D. degree in Electrical and Computer Engineering from The University of Texas at Austin in 2016, where he was advised by Prof. Andrea Alù. Dr. Monticone joined the faculty of Cornell University in January 2017.

Dr. Monticone has authored and co-authored more than 100 scientific contributions in peer-reviewed journal papers, book chapters, and peer-reviewed conference proceedings, and has given over 20 invited talks and seminars. Dr. Monticone is serving as a reviewer for dozens of journals, international conferences, and funding agencies. He has organized and chaired various special sessions in international symposia and conferences. He has been a member of the organizing committee of the Metamaterials congress series since 2015. Dr. Monticone's current research interests are in the areas of applied electromagnetics, engineered metamaterials and metasurfaces, and theoretical/computational nanophotonics, with particular focus on innovative and extreme aspects of wave interaction with engineered materials and nano-structures.

Over the last few years, he has received several research awards, including the *Leopold B. Felsen Award for Excellence in Electrodynamics* from the European Association on Antennas and Propagation (2019), the *AFOSR Young Investigator Program Award (YIP)* from the U.S. Air Force Office of Scientific Research (2018), the *Inaugural Margarida Jacome Dissertation Award* from The University of Texas at Austin (2017), the *Raj Mittra Travel Grant Award* (2017), the *WNCG Student Leadership Award* (2016), and *Doctoral Research Awards* from both the *IEEE Antennas and Propagation Society* (2013) and the *IEEE Photonics Society* (2015). Dr. Monticone is a member of the IEEE, the American Physical Society (APS), the Optical Society of America (OSA), The International Society for Optics and Photonics (SPIE) and is a full member of the International Union of Radio Science (URSI).

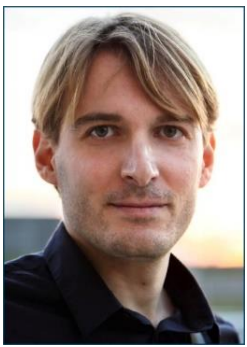

**Andrea Alù** received the Laurea, M.S., and Ph.D. degrees from the University of Roma Tre, Rome, Italy, in 2001, 2003, and 2007, respectively. Andrea Alù is the founding director of the Photonics Initiative at the CUNY Advanced Science Research Center, Einstein Professor of Physics at the CUNY Graduate Center, and Professor of Electrical Engineering at The City College of New York. His research interests span over a broad range of technical areas, including applied electromagnetics, nano-optics and nanophotonics, microwave, THz, infrared, optical and acoustic metamaterials and metasurfaces, plasmonics, nonlinearities and nonreciprocity, cloaking and scattering, acoustics, optical nanocircuits, and nanoantennas. He is a Fellow of IEEE, OSA, APS and SPIE, and a recipient of several awards, including the 2015 NSF Alan T. Waterman Award, the 2019 IEEE Kiyo Tomiyasu Award, the 2013 OSA Adolph Lomb Medal and the 2011 Issac Koga Gold Medal.